\documentclass{aa}
\usepackage{hyperref}
\hypersetup{
    unicode=false,          % non-Latin characters in Acrobat?s bookmarks
    colorlinks=true,       % false: boxed links; true: colored links
    linkcolor=blue,          % color of internal links (change box color with linkbordercolor)
    citecolor=blue,        % color of links to bibliography
    filecolor=blue,      % color of file links
    urlcolor=blue           % color of external links
}

\usepackage{threeparttable}
\usepackage{graphicx}
\usepackage{adjustbox}
\usepackage{txfonts}
\usepackage{lipsum}
\usepackage{booktabs}
\usepackage{xcolor}

\usepackage{subcaption}         % necessary for continued figures, example in section 3
\usepackage{lscape}             % to rotate a single page table, example in appendix.
\usepackage{placeins}           % useful with \FloatBarrier, to keep 
\usepackage[version=4]{mhchem}

\newcounter{maacounter}
\newenvironment{maaenvironment}[1][]{\refstepcounter{maacounter} \nobreakspace 
{\color{orange}{MAA(\themaacounter):~{#1}}} \rmfamily}{}

\usepackage[normalem]{ulem} % keeps \emph behavior

\newcommand{\Aenus}{\texttt{Aenus-ALCAR}\xspace}
\newcommand{\GENEC}{\texttt{GENEC}\xspace}
\newcommand{\MESA}{\texttt{MESA}\xspace}
\newcommand{\ag}{\texttt{G20}\xspace}
\newcommand{\ad}{\texttt{M13}\xspace}

\defcitealias{Griffiths2026PaperI}{Paper I}
\defcitealias{Griffiths2026PaperII}{Paper II}

\let\linenumbers\relax
\let\nolinenumbers\relax
\begin{document}

%%%%%%%%%%%%%%%%%%%%%%%%%%%%%%%%%%%%%%%%
% if you use custom commands in your title,
% ensure to check your title when submitting!
%%%%%%%%%%%%%%%%%%%%%%%%%%%%%%%%%%%%%%%%
   \title{The first 3D MHD core-collapse progenitors II:} 
%MAA: preferred subtitle
%   \subtitle{Rotation, magnetic field geometry and amplification.} 
   \subtitle{Rotation, magnetic-field amplification, and magnetic topology} 

%%%%%%%%%%%%%%%%%%%%%%%%%%%%%%%%%%%%%%%%
% Please separate each author with the \and command
%
% Please do not include ORCIDs next to author names.
% Only ORCIDs authenticated by individual authors in EDPS
% editorial system will be taken into account.
% ORCIDs included here will be removed.
%%%%%%%%%%%%%%%%%%%%%%%%%%%%%%%%%%%%%%%%

   \author{A. Griffiths\inst{1,2}
        \and Miguel-{\'A}. Aloy\inst{1,3}
        \and M. Obergaulinger\inst{1,3}
        }

 \institute{Departament d'Astronomia i Astrofísica, Universitat de València, 46100 Burjassot, Spain
 \and
Astrophysics Group, Lennard-Jones Laboratories, Keele University, Keele ST5 5BG, UK
\\
 \email{a.griffiths@keele.ac.uk, miguel.a.aloy@uv.es}
\and
Observatori Astronòmic, Universitat de València, 46980 Paterna, Spain}
\date{Received September 30, 20XX}
% \abstract{}{}{}{}{}
% 5 {} token are mandatory
 
  \abstract
   {The most energetic core-collapse supernovae are thought to arise from rapidly rotating, magnetised progenitors. However, the three-dimensional pre-collapse structure of their angular momentum and magnetic fields remains poorly constrained, limiting the realism of magnetorotational core-collapse simulations.} 
   {We investigate the angular-momentum distribution, magnetic-field amplification and magnetic topology of physically consistent three-dimensional magnetohydrodynamic pre-supernova progenitors. This second paper focuses on the rotational and magnetic properties of the models introduced in \citetalias{Griffiths2026PaperI}.} 
   {We used \Aenus to evolve two compact Wolf--Rayet progenitors, computed with the stellar-evolution codes \GENEC and \MESA, through the final minutes before core collapse. The initial magnetic-field strengths were derived from the one-dimensional stellar-evolution prescriptions in radiative regions and then evolved self-consistently in three dimensions.}
   {Our models suggest that the rotation profile near the inner core can depart from a purely shellular distribution and reorganise toward a more cylindrical structure.
   %MAA: previous sentence softens the point.
   %We find that the angular-momentum distribution near the inner core is better represented by a cylindrical rotation law than a shellular one. 
   In convective regions, hydrodynamic Reynolds stresses drive the flow toward an approximately constant specific-angular-momentum profile, corresponding to an average rotation profile close to $\Omega\propto \varpi^{-2}$ ($\varpi$ denotes the cylindrical radius). Maxwell stresses oppose this tendency but are not strong enough in our models to restore rigid rotation. Convective regions amplify seed magnetic fields, transported from neighbouring radiative layers, producing saturated fields with comparable toroidal and poloidal components and a topology containing substantial small-scale power. As a result, regions that are magnetically disconnected in the original one-dimensional stellar-evolution description become magnetically linked in the multidimensional models. Together with \citetalias{Griffiths2026PaperI}, these models represent the first 3D MHD pre-supernova progenitors of this kind, suitable for subsequent collapse and explosion calculations.}
   {Multidimensional evolution can substantially modify both the angular-momentum distribution and magnetic topology of pre-collapse progenitors. They provide a physically motivated basis for constructing more realistic initial conditions for magnetorotational core-collapse simulations and for improving prescriptions of magneto-convective angular-momentum transport in late stellar evolution.
}

%MAA: too many keworkds


\maketitle

%%%%%%%%%%%%%%%%%%%%%%%%%%%%%%%%%%%%%%%%%%%%%%%%%%%%%%%%%%%%%%
\section{Introduction}

%%%%%%%%%%%%%%%%%%%%%%%%%%%%%%%%%%%%%%%%%%%%%%%%%%%%%%%%%%%%%%

Massive stars with rapid rotation and strong magnetic fields are leading candidates for the progenitors of the most energetic core-collapse supernova (CCSNe), including hypernovae and some long gamma-ray bursts \citep[e.g.][]{Woosley_Bloom_2006,Nomoto_Tanaka_Tominaga_Maeda_Mazzali_2007,Burrows_2007,Mueller_2024arXiv240318952}. Magnetorotational explosions have also been proposed as favourable sites for the production of heavy $r$-process nuclei \citep[e.g.][]{Reichert_2022,Zha_muller_Powell_2024}, and their nucleosynthetic yields can depend sensitively on the strength and topology of the progenitor magnetic field \citep{Reichert_2024}. The angular-momentum (AM) and magnetic-field distributions in the stellar core are therefore central ingredients for predicting the explosion geometry, the compact-remnant spin, and the magnetic-field structure inherited by the nascent neutron star.

Current magnetorotational CCSN calculations are usually initialised from one-dimensional stellar-evolution (SE) models \citep[e.g.][]{Mosta_Richers_Ott_Haas_Piro_Boydstun_Abdikamalov_Reisswig_Schnetter_2014,Obergaulinger2020}. Such models can only estimate magnetic-field strengths in radiative regions through effective prescriptions for magnetic instabilities, such as the Tayler--Spruit (TS) dynamo \citep{Spruit_2002} or, in some cases, the magneto-rotational instability \citep[MRI;][]{Balbus_Hawley_1991}. However, they do not provide the global three-dimensional (3D) geometry of the field, nor the magnetic connectivity amongst the iron core, radiative layers, and convective shells. This limitation is particularly important because multidimensional simulations of magnetorotational explosions have shown that the field topology, not only its amplitude, can influence the explosion dynamics and outcome \citep[e.g.][]{Obergaulinger_2017MNRAS.469L..43, Aloy_2021,Bugli_Guilet_Obergaulinger_2021}.

Existing multidimensional MHD studies have begun to clarify the role of convective burning shells in magnetic-field amplification and AM transport \citep[e.g.][]{Varma_2021,Varma_2023}. However, they do not yet provide whole-star, self-consistent, rotating and magnetised pre-collapse progenitors extending from the iron core to the outer stellar layers--in particular, excising the core precludes determining how the magnetic flux connects the inner regions to the overlying convective shells, and how the resulting field topology would be inherited by the collapsing core.

Beyond their use as initial conditions for CCSNe simulations, multidimensional MHD progenitor models can also provide feedback for SE modelling. Stellar-evolution calculations necessarily rely on effective one-dimensional prescriptions for convection, AM transport, and magnetic stresses. The companion paper of this series \citep[][hereafter \citetalias{Griffiths2026PaperI}]{Griffiths2026PaperI} focused on convection and nuclear burning in the final pre-collapse shells. In the present paper (\citetalias{Griffiths2026PaperII}, hereafter) we address the complementary problem: the evolution of rotation, magnetic-field amplification, and magnetic topology during the final minutes before collapse. 
Related work on magnetised stellar interiors has often focused on lower-mass or solar-type stars \citep{Jouve_Gastine2015,Emeriau-Viard_Brun_2017,Gouhier_Lignieres_Jouve_2021}, while recent studies have begun to explore how multidimensional magneto-convective simulations can inform one-dimensional prescriptions for massive stars \citep{Shimada_McNeill_Varma_Maeda_Yokoyama_Muller_2026}.

The models analysed here are the first 3D MHD pre-supernova progenitors of this kind, including both the iron core and the surrounding burning shells. Their global properties, numerical setup, hydrostatic initialisation, turbulent shell structure, and nuclear-burning behaviour are presented in \citetalias{Griffiths2026PaperI}. Here we focus on the rotational and magnetic properties of the same models. We examine how the AM distribution evolves in convective and radiative regions, how magnetic fields are amplified in shells where SE models predict no fields, and how the final magnetic topology differs between radiative and turbulent layers.

This paper is structured as follows. Section~\ref{sec:ini_3D} summarises the magnetic and rotation history of the progenitor models and describes how we initialise multidimensional models from the one-dimensional SE information. Section~\ref{sec:results} presents the evolution of AM, magnetic-field amplification, and magnetic topology during the final pre-collapse phase. We then discuss in Section~\ref{sec:discussion} the implications of our models and how they may be used to improve rotation and magnetic modelling in pre-supernova progenitors. Finally, Section~\ref{sec:conclusions} summarises our main conclusions.

%\textcolor{blue}{AG: Shorter introduction here and taking a lot from the intro of Paper I. I wonder if magnetars and pulsars may be mentioned more generically in this introdcution as we are deling with rotation and magnetic fields. I have no strong opinions if you feel the science content is sufficent then great. If not let me know and I will develop on a few more key points.}
%\maaC{Not in my opinion. The magnetar field is most likely set by amplification in the PNS. Certainly, starting from larger fields could help, but this is only part of the story to get suitable fields.}

%MAA: this is my proposal for a new title
\section{Rotational and magnetic initial conditions}

%\section{Initialisation of 3D models \textcolor{blue}{AG:Proably want a better title}}

\label{sec:ini_3D}
The SE progenitors and the multidimensional numerical setup are described in detail in \citetalias{Griffiths2026PaperI}. We consider two compact Wolf--Rayet-like progenitors. The first, \ag, is a $20\,M_\odot$ model computed with the version of GENEC presented in \cite{Griffiths_2025}. The second, \ad, is taken from series B of \citet{Aguilera-Dena_2020}, computed with \MESA. In the present paper we summarise only the aspects of the one-dimensional models that are directly relevant for the rotational and magnetic-field structure of the subsequent 3D MHD simulations.

\subsection{Rotational and magnetic properties of 1D models}

Models \ag and \ad are both fast rotators at birth, with initial surface rotation rates equal to $460 \rm \, km \, s^{-1}$ and $600 \rm \, km \, s^{-1}$ respectively. Transport of AM is induced through  the TS-dynamo during SE. For model \ag, the magnetic transport follows the prescription of \cite{Eggenberger_2022} whereas \ad uses the description of \cite{Heger_Woosley_Spruit_2005}--the two only differ by a calibration factor included in \cite{Eggenberger_2022}.
%MAA: already said in the intro of the section 2.
%Further details of the progenitor setup are given in \citetalias{Griffiths2026PaperI}.

The fast initial rotation combined with the strong coupling of rotation induced by magnetic torques results in a large amount of mass and AM loss driven by the fast velocities at the surface of each star. These winds result in the loss of the hydrogen envelope of both stars.\footnote{Both models end as compact, hydrogen-poor progenitors. However, the origin of this structure differs between them: while \ag is stripped of its hydrogen-rich envelope, \ad becomes hydrogen- and helium-depleted through quasi-chemically homogeneous evolution, where rotational mixing and nuclear burning act together with rotationally enhanced mass loss.}

\begin{table*}[ht]
\centering
\caption{Total mass, total AM, and surface rotation of the progenitor models at different evolutionary phases. We define the TAMS as the point where central hydrogen mass fraction drops below $10^{-5}$. Helium depletion is operatively defined as the time when the central helium mass fraction drops below $10^{-5}$. The ``final model'' denotes the last model computed in the SE calculation.}
\small
\begin{tabular}{ c c c c c c c c c c c c }
\hline\hline
 & \multicolumn{3}{c}{ZAMS} & \multicolumn{3}{c}{TAMS} &  \multicolumn{3}{c}{He depletion} & \multicolumn{2}{c}{Final model} \\
\cline{2-3}\cline{5-6}\cline{8-9}\cline{11-12}

& \ag & \ad & & \ag & \ad & & \ag & \ad & & \ag & \ad \\

$M_{\rm tot} \ [M_{\odot}]$ & 19.99 & 12.99 & & 17.41 & 12.47 & & 15.19 & 10.92 & & 15.15 & 10.37\\
$L_{\rm tot} \ [10^{52} \rm \ g \ cm^2 \ s^{-1}]$ & 4.74 & 2.85 && 1.33 & 2.00 & & 0.0118 & 0.516 & & 0.0113 & 0.243 \\
$\Omega_{\rm surf} \ [10^{-4} \rm \ rad \ s^{-1}]$ & 1.30 & 2.31 && 1.58 & 4.38 & & 0.192 & 18.2 & & 0.942 & 43.8 \\
\hline
\end{tabular}
\label{tab:AM_evol}
\end{table*}

Table~\ref{tab:AM_evol} summarises the total mass, total AM, and surface rotation rates at key evolutionary phases. By the end of core-helium burning, model \ag retains only a total AM of $\simeq 10^{50}\, \rm g \, cm^2 \, s^{-1}$, compared to $\simeq 5 \times 10^{51}\, \rm g \, cm^2 \, s^{-1}$ for model \ad. These values correspond to approximately 1\% and 25\%, respectively, of their total AM at terminal age main sequence (TAMS). In both models, the greatest AM loss occurs after the TAMS, but this loss is much larger for \ag. Thus, although the two progenitors have broadly similar compact pre-collapse structures (see Fig. 1 of \citetalias{Griffiths2026PaperI}), they differ strongly in their rotational content: \ad is much faster rotating than \ag both at the surface and in the core at the time of mapping to 3D.

The magnetic-transport prescriptions provide estimates of the saturated magnetic-field strengths in radiative regions, the exact expressions are given in Appendix~\ref{sec:appen_field}.  Figure~\ref{fig:B_and_O} shows the toroidal and poloidal saturation-field strengths used to initialise the 3D MHD-models. Convective regions are highlighted in light blue; in the one-dimensional SE models, these regions are not assigned initial magnetic-field strengths.

The field strengths obtained from these prescriptions should be regarded as order-of-magnitude estimates. In some layers, the magnetic energy implied by the saturation formulae, $ E_{\rm B}\sim \int_{\rm Shell}\frac{1}{2}B^2 dV$, can become comparable to, or even exceed, the local rotational energy $  E_{\Omega} \sim \int_{\rm Shell}\frac{1}{2}\rho (r\sin \theta \Omega)^2 dV$, where $B$ and $\Omega$ are the local magnetic field strength and rotational frequency, respectively.
This would be unphysical if interpreted literally, since the magnetic energy ultimately derives from  differential rotation. In such cases, a back-reaction should limit further field amplification. To prevent this situation in the initial conditions, we cap the toroidal saturation field so that the local ratio of magnetic to rotational energy does not exceed 0.1.

\begin{figure}[t!]
    \centering
    \includegraphics[width=\columnwidth]{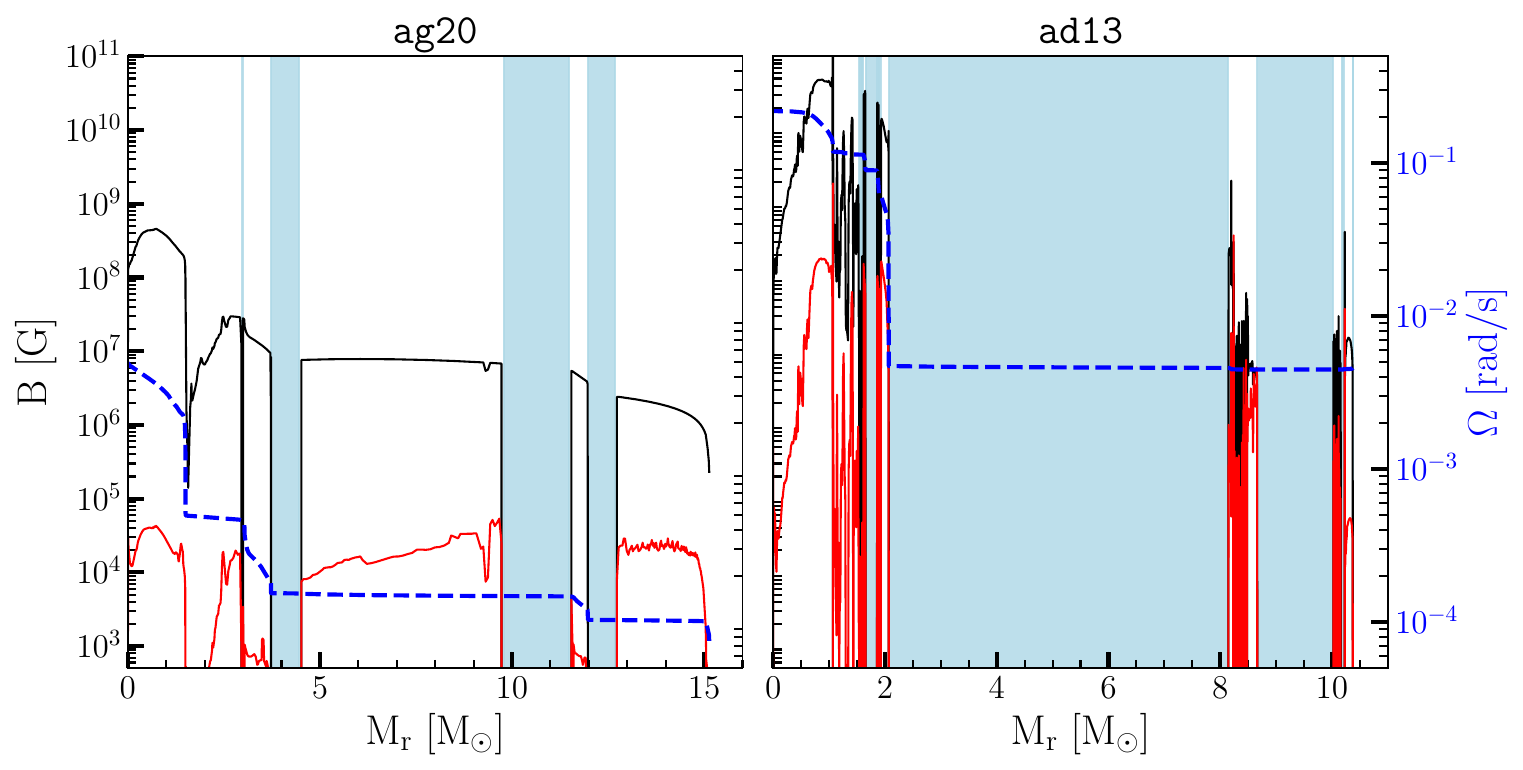}
    \caption{Saturation field strengths at mapping for the toroidal component (black) and poloidal component (red) associated with the magnetic instabilities considered in model \ag (left) and \ad (right). The light blue bands indicate convective regions, where 1D SE models do not provide magnetic-field estimates. The blue dashed line shows the rotational frequency $\Omega$.}%
    \label{fig:B_and_O}%
\end{figure}

The peak field strengths in \ag are at least two orders of magnitude lower than in \ad (Fig.~\ref{fig:B_and_O}), primarily because of the much slower core rotation in \ag. The estimated poloidal field is likewise weaker in \ag, only reaching $10^5$\,G, whereas  it peaks at $\sim 10^8$\,G in the core of \ad.  If the 1D magnetic structure were taken at face value, the magnetised core would be separated from the outer magnetised layers by convective shells with no prescribed magnetic field (blue regions in  Fig.~\ref{fig:B_and_O}). One of the central results of the present paper is that the 3D MHD evolution alters this picture: magnetic flux is transported into convective regions and amplified there, leading to magnetic connectivity across layers that are disconnected in the original SE estimate (Section~\ref{sec:results}).

\begin{figure}[t!]
    \centering
    \includegraphics[width=0.85\columnwidth]{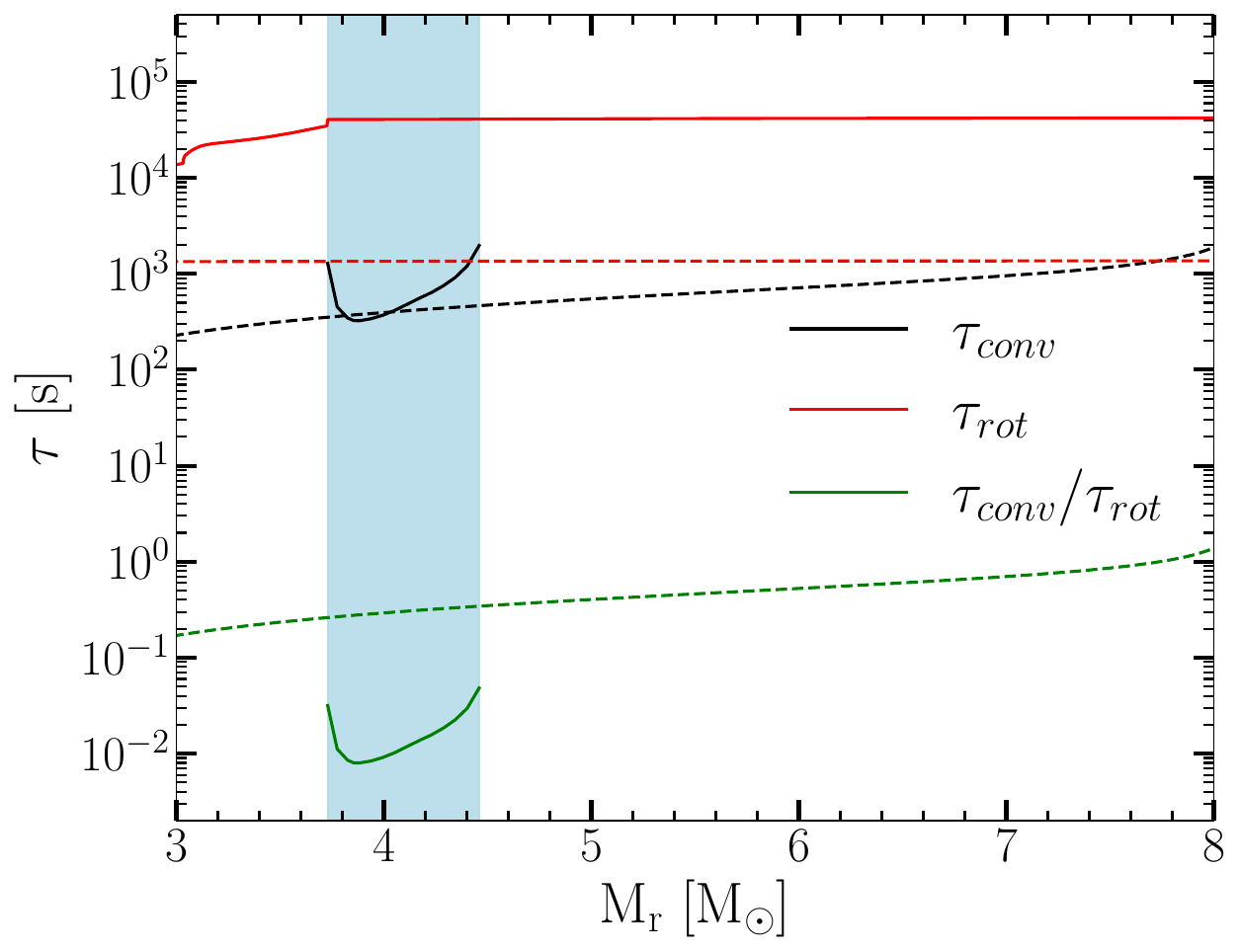}
    \caption{Comparison of the convective turnover time $\tau_{\rm conv}$ (black), rotational timescale $\tau_{\rm rot}$ (red), and their ratio (green) for models \ag (solid) and \ad (dashed) at the mapping point. The blue-shaded region marks the oxygen convective shell of \ag; the plotted range lies fully inside the broader oxygen convective shell of \ad.}%
    \label{fig:conv_vs_omega}%
\end{figure}

The amplification of magnetic fields in stellar convective regions depends on the relative importance of convection and rotation. To assess this, we compare the corresponding  convective turnover time and local rotational period,\footnote{We denote these timescales with ``SE'' as they are computed using values from the SE models. This is in contrast with similar quantities that we will compute in later sections based on the 3D simulations.}
\begin{equation}
    \tau_{\rm conv, SE} = \frac{2 H_p}{v_{\rm conv}},
    \label{eq:tau_conv}
\end{equation}
where $v_{\rm conv}$ is the MLT estimate of the convective velocity and $H_p$ is the pressure scale height. We estimate the rotational timescale with the local rotational period, 
\begin{equation}
    \tau_{\rm rot, SE} =  \frac{2 \pi}{\Omega}.
\end{equation}
The ratio of these two timescales is related to the Rossby number \citep[see][for a similar, yet non-identical, definition]{Varma_Mueller_2023} and measures whether rotation or convection dominates the local dynamics.

Figure~\ref{fig:conv_vs_omega} shows these timescales along with their ratio in the convective oxygen-burning shells for each progenitor (see the shaded regions of Fig.~\ref{fig:B_and_O}).
%\footnote{For comparison between progenitors we show the oxygen shell, however, the silicon burning shell of \ad harbours the fastest rotation compared to the local convective turnover time.}
%
In both models, the ratio is smaller than unity--i.e., convection is faster than rotation, as expected, since convection in advanced burning phases is extremely rapid and AM transport mediated by magnetic fields has substantially slowed the stellar cores. The ratio is roughly an order of magnitude larger in \ad than in \ag, reflecting the much faster rotation of the \MESA progenitor. 

\subsection{Initial magnetic-field and angular-momentum distributions.}

The effective implementation of rotation and magnetic fields in 1D SE cannot specify their full 3D structure. Mapping such models into multidimensional MHD calculations therefore requires assumptions about both the angular-velocity distribution and the magnetic-field topology.

In 1D SE, rotation is usually assumed to be shellular, $\Omega = \Omega(r)$. This approximation is motivated by efficient horizontal transport along isobars, which tends to homogenise the rotational frequency on spherical shells during most evolutionary phases \citep[e.g.][]{Maeder_2003}. Near core collapse, however, the relevant dynamical and convective timescales become short, and the rotation profile need not remain shellular.
%MAA: joined paragrphs
%
An alternative limiting case is cylindrical rotation, in which $\Omega$ is constant on coaxial cylinders, i.e., $\Omega=\Omega(\varpi)$, where $\varpi=r\sin\theta$ is the distance to the rotational axis. 
%\delt{With this description, matter in the core could rotate at the same speed as material in the outer layers.}\maaC{so far this extra information does not provide a clear insight. Maybe later we can use it in the argumentation.} 

In the deep stellar interior, as the core approaches collapse, the true rotational structure likely lies between these two idealised limits and may vary between radiative and convective regions. Multidimensional simulations of rotating stars indeed show a reorganisation of rotation towards cylindrical isosurfaces within the envelope \citep{Charbonneau_MacGregor_1993, Browning_Brun_Toomre_2004}. In radiative zones, the preferred differential-rotation structure can also depend on the Prandtl number\footnote{This is the ratio of kinematic viscosity to thermal diffusivity, which is generally very low in stellar interiors ($\sim 10^{-6}$).}  \citep{Gouhier_Lignieres_Jouve_2021}. 

\begin{figure}[t!]
    \centering
    \includegraphics[width=\columnwidth]{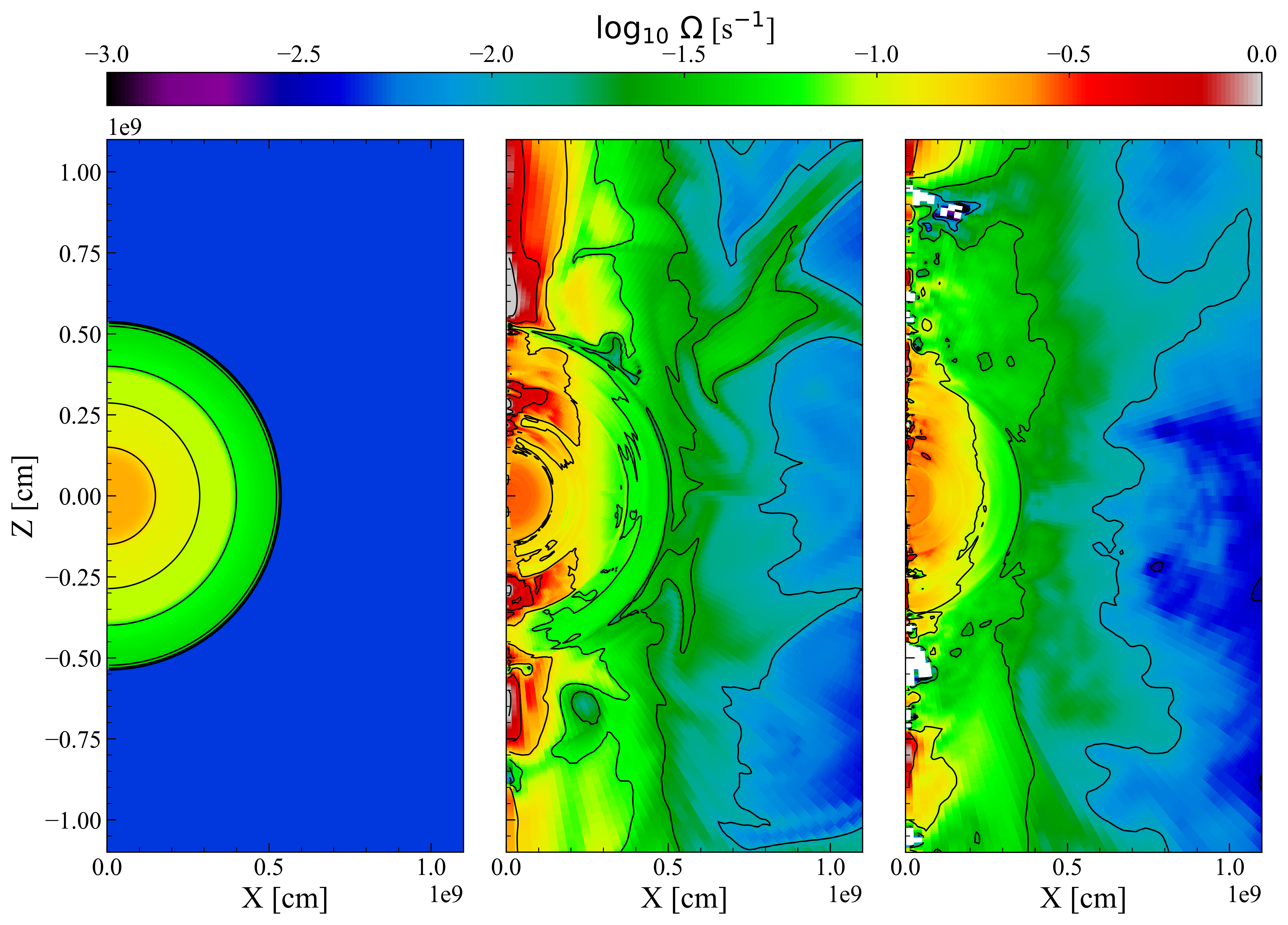}
    \caption{Two-dimensional maps of the rotational frequency for the initial setup of a 2D test with shellular rotation (left), the resulting profile after 400\,s of evolution (centre), and the $\phi$-averaged 3D profile after 400\,s of total evolution (right), obtained from a model initialised with the cylindrically weighted rotation profile described in the text. Black lines indicate iso-contours of $\Omega$.}%
    \label{fig:omega_spherical_vs_2d}%
\end{figure}

To assess the rotational profile in our progenitors, we performed a two-dimensional test simulation of model \ad.
Figure~\ref{fig:omega_spherical_vs_2d} compares the initial shellular angular-velocity distribution (left) with the state after 400\,s of evolution (centre). The model rapidly develops a more cylindrical organisation outside the inner core, while the central region ($r \leq 10^8 \, \rm cm$) remains closer to shellular rotation. 
To reduce the initial transient associated with this readjustment, the 3D models are initialised with a cylindrically weighted angular-velocity distribution ($\Omega(r,\theta) = \Omega_{\rm SE} (r) /  ( \sin \theta )$) rather than with a purely shellular one.
%
%\footnote{\rplc{To prevent $\Omega ( r, \theta)$ from becoming excessively large near the axis, we cap its value at}{In the implementation used here, the mapped rotational frequency is capped near the rotation axis to avoid unphysically large values. The precise} mapping \insr{should be kept consistent with the simulation setup, }so that \rplc{it}{outside the core the rotational frequency} does not exceed the core rotational rate.
%}  
%MAA: something was broken if I did my normal correction/edition in this footnote. I made a simpler one below.
\footnote{
In the implementation used here, the mapped rotational frequency is capped near the rotation axis to avoid unphysically large values. The precise mapping should be kept consistent with the simulation setup, so that outside the core the rotational frequency does not exceed the core rotational rate.
}

This choice of initial setup was motivated by the 2D axisymmetric test. To check whether this cylindrical-like organisation is preserved during the subsequent 3D evolution, the right panel of Fig.~\ref{fig:omega_spherical_vs_2d} shows the $\phi$-averaged angular-velocity map of the 3D model at $t=400\,$s, starting from the cylindrically weighted initial rotation profile. The resulting profile still shows a clear departure from shellular rotation, indicating that the cylindrical-like structure is not erased by the fully 3D turbulent evolution.
%

%This choice  places the model closer to the rotational configuration preferred by the multidimensional evolution.
%MAA: full stop

%\delt{The actual rotational structure will naturally depend on the specific model (e.g., on the rotation rate) as well as on the location within the star, with different sized shells possibly favouring different geometries. Nonetheless, the evidence presented here suggests that mapping pre-SN models assuming a shellular distribution is likely to misrepresent the true rotational structure at the onset of collapse.}\maaC{This is more for a discussion/conclusions section that for here.}

%\begin{figure}[t!]
 %   \centering
%    \includegraphics[width=0.85\columnwidth]{figures/omega_spherical_vs_2d.png}
%    \caption{Two-dimensional map of the rotational frequency for an axisymmetric test run of model \ad. Left panel shows $\Omega$ initialised with a shellular rotation profile, while the  right panel displays the configuration after 300\,s of evolution. Black lines indicate iso-contours of $\Omega$.}%
%    \label{fig:omega_spherical_vs_2d}%
%\end{figure}

The magnetic-field strength can be estimated from the 1D SE prescriptions in radiative regions, but the field geometry is not specified. We therefore construct an initial axisymmetric field using the SE toroidal and poloidal saturation strengths. The toroidal component is assigned as
\begin{align}
    B_{\rm tor} (r,\theta,\phi ) & = b_{\rm tor} ( r ), 
\end{align}
where $b_{\rm tor}$ is the SE toroidal saturation field (see Appendix~\ref{sec:appen_field}). The radial component is written as
\begin{align}
    B_{r} (r,\theta,\phi ) & = b_{\rm pol} ( r ) \cos ( n \theta ),
\end{align}
where $b_{\rm pol}$ is the SE poloidal saturation field and we set $n=1$ to generate a dipolar structure in each radial shell. The polar component $B_{\theta} ( r,\theta,\phi)$ is then obtained by imposing the solenoidal constraint 
\begin{equation}
    \nabla \cdot B = 0
\end{equation}
thus ensuring that the constructed initial field is divergence-free. 
%MAA: full stop

\begin{figure}[t!]
    \centering
    \includegraphics[width=0.7\columnwidth]{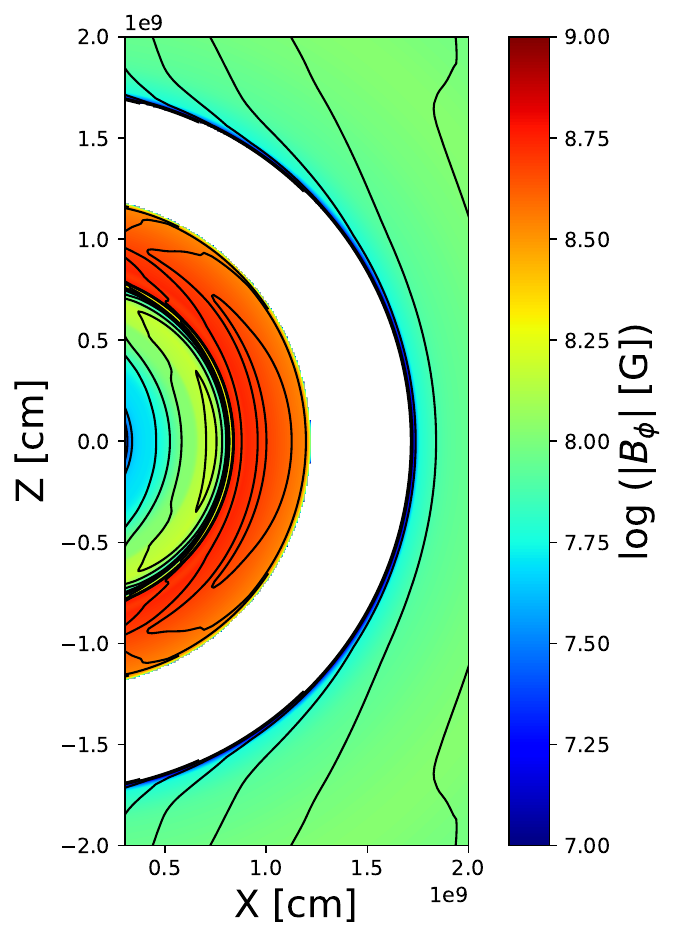}
    \caption{Axial projection of initial poloidal magnetic-field lines for model \ag. The background colour map shows the toroidal magnetic-field strength. The white region corresponds to layers where no magnetic field is prescribed by the 1D SE model.}%
    \label{fig:face_on_magnetic_field_streamlines}%
\end{figure}

Figure~\ref{fig:face_on_magnetic_field_streamlines} shows the resulting initial field geometry for model \ag. The background colour map gives the toroidal field strength, while the streamlines represent the poloidal field. White regions show where the SE model does not initially prescribe a magnetic field, namely convective layers.

\section{Results}
\label{sec:results}

Using the setup outlined in Section~\ref{sec:ini_3D} and in \citetalias{Griffiths2026PaperI}, we evolved both progenitors through the final minutes prior to collapse. Model \ag was evolved for 190\,s in 3D, while model \ad was run for 320\,s in 3D, after an initial 200\,s of axisymmetric relaxation phase. As discussed in \citetalias{Griffiths2026PaperI}, these durations cover many convective turnovers in the inner burning shells, although the outer convective regions are less well sampled. The resulting models therefore provide a statistically robust description of the rotational and magnetic evolution in the inner convective regions, while the outer shells should be interpreted more cautiously.

The multidimensional evolution modifies  both the AM distribution and the magnetic-field structure relative to 1D SE predictions in the convective regions. In the following, we first analyse the AM redistribution driven by hydrodynamic and magnetic stresses. We then examine magnetic-field amplification in radiative and convective regions, and finally characterise the topology of the magnetic field at the pre-SN link.

For reference, the region labels R1, C1, R2 and C2 follow the notation of \citetalias{Griffiths2026PaperI}: R1 denotes the radiative iron-core region, C1 the innermost turbulent shell, R2 the intervening radiative layer, and C2 the outer oxygen-burning convective shell. In model \ad, C1 corresponds to a thin Si-burning shell, whereas in \ag it corresponds to the turbulent edge of the iron core.

\subsection{Angular momentum evolution}

\begin{figure}[t!]
    \centering
    \includegraphics[width=\columnwidth]{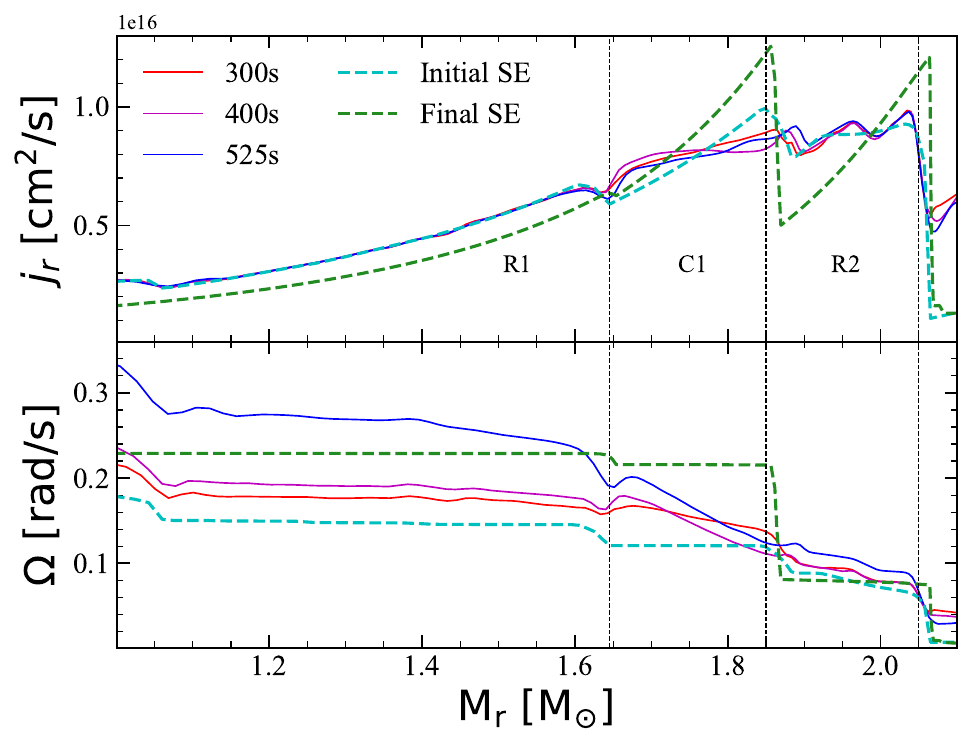}
    \caption{Shell-averaged specific AM, top, and rotational frequency, bottom, for model \ad at different evolutionary times. The cyan and green lines show the corresponding SE profiles at mapping and at the pre-SN link, respectively. The figure focuses on the inner convective region C1 and its neighbouring radiative layers; the outer parts of the star are not shown.  
    }%
    \label{fig:jr_omega}%
\end{figure}

%\maaC{The vertical range of the bottom panel could be enlarged to show more clearly the difference in $\Omega$ of the 3D model compared to the final SE profile in region R1. Maybe that makes the differences elsewhere not so clearly visible, but, as represented, one does not grasp how large the change between 400\,s and 525\,s is.}
%\maaC{I wonder how large is the scattering of $j_r$ about the mean, specially in the convective region. Maybe one could make a figure (for now just internal) showing the shell-averaged $j_r$ and with light-gray colour, the $\phi$-averaged $j_{r}(r,\theta)$ for all possible values of $\theta$. What I aim to assess is whether the different trends for $j_r$ (constant or growing as $r^2$) are consistent with the spread seen in 3D MHD models. }
%Will do this in the coming days.

%\maaC{A bit of reorganization in the following paragraphs to discuss first what happens in raditive regions and then in convective layers. I think this way shows more clearly the impact of the evolution in the convective layers.}
We define the specific angular momentum about the rotation axis as
$j = \Omega \varpi^2$, where $\varpi=r\sin\theta$  is the cylindrical radius.
In radiative regions, the shell-averaged specific AM remains close to the initial SE value during the short pre-collapse interval considered here (Fig.~\ref{fig:jr_omega}). The radiative region R1 spins up as the star contracts, but its specific angular momentum is approximately conserved. The neighbouring radiative shell R2 shows small fluctuations in the multidimensional profiles, but its average evolution remains close to the SE prediction. 

The main difference between the 3D MHD models and the SE predictions occurs in the convective region C1 (see the case of model \ad in Figure~\ref{fig:jr_omega}).
In the SE model, this shell evolves toward nearly uniform rotational frequency. In the 3D model, by contrast, the shell develops substantial radial differential rotation: the base of the shell rotating faster than the top. 
Equivalently, the convective region evolves toward an approximately constant specific-angular-momentum profile. Since $j\simeq \mathrm{const.}$ implies $\Omega\propto \varpi^{-2}$, the resulting rotation profile differs markedly from the $\Omega\simeq\mathrm{const.}$ behaviour usually assumed for convective regions in 1D SE calculations.

A similar tendency is present in the convective regions of model \ag, although it is harder to quantify there. Because \ag rotates  much more slowly than \ad, the azimuthal velocity associated with turbulent convective motions becomes comparable to, or even larger than, the mean rotational velocity in that model. For this reason, we occasionally observe negative values of local or shell-averaged $\Omega$ and $j_r$ when convective motions dominate the azimuthal flow. This makes the interpretation of AM redistribution in \ag less clean than in the more rapidly rotating model \ad.

In the outer convective shell C2 of \ad, the same qualitative trend is visible: the base of the shell begins to spin up relative to the top. However, as shown in \citetalias{Griffiths2026PaperI}, this outer shell undergoes fewer convective turnovers during the 3D evolution. Its AM distribution has therefore not reached a quasi-stationary state as well defined as that of C1. We consequently focus the quantitative analysis below on the inner convective shell C1 of \ad.

The tendency toward a constant-$j_r$ is consistent with hydrodynamic simulations of \cite{Yoshida_2021} and \cite{Varma_Mueller_2023}, where turbulent motions in convective shells redistribute angular momentum away from the $\Omega=\mathrm{const.}$ state predicted by SE models. However, \cite{Varma_Mueller_2023} found in their MHD model that magnetic stresses eventually counteracted the Reynolds stresses and restored a flatter angular-velocity profile. Since our models also develop magnetic fields in the convective regions it is necessary to determine why the same reversal does not occur here.

To analyse the AM transport, following \cite{Varma_Mueller_2023}, we write the radial part of the AM conservation equation\footnote{Equation~14 in \cite{Varma_Mueller_2023} contains typographical errors: the sign in front of the spatial derivative term $\nabla_r$ should be positive and the sign in front of the magnetic term negative. As we use Heaviside-Lorentz units,  the $4\pi$ factor is not included in our definition.} as,
\begin{equation}
\label{eq:AM_transport}
    \frac{\partial \langle \rho v_{\phi} r \sin \theta \rangle }{\partial t} = - \nabla_r \cdot \langle\rho v_r v_{\phi}r\sin \theta - B_r B_{\phi} r\sin\theta\rangle,
\end{equation}
where $\nabla_r$ denotes the radial term of the divergence and the angular brackets indicate a shell average. Applying the Reynolds and Favre decompositions \citep{Favre_1965} defined in Appendix~\ref{sec:appen_favre}, the radial AM flux can be decomposed into the following terms,
\begin{align}
\label{eqn:fluxes_adv}
   \rm Advective \ :&  \  \langle \rho \widetilde{v_r} ( \widetilde{\Omega} + \Omega'') r^2 \sin^2 \theta\rangle, \\
\label{eqn:fluxes_mer}
   \rm Meridional\  :&  \ \langle\rho v_r'' \widetilde{\Omega} r^2 \sin^2 \theta\rangle, \\ 
\label{eqn:fluxes_aturb}
   \rm Turbulent \ :& \ \langle\rho v_r'' \Omega'' r^2 \sin^2 \theta\rangle, \\
\label{eqn:fluxes_mag}
\rm Magnetic \ :&  \  -\langle B_r B_{\phi} r \sin \theta \rangle.
\end{align}
The first three terms are Reynolds fluxes, while the last term is the Maxwell flux.

\begin{figure}[t!]
    \centering
    \includegraphics[width=\columnwidth]{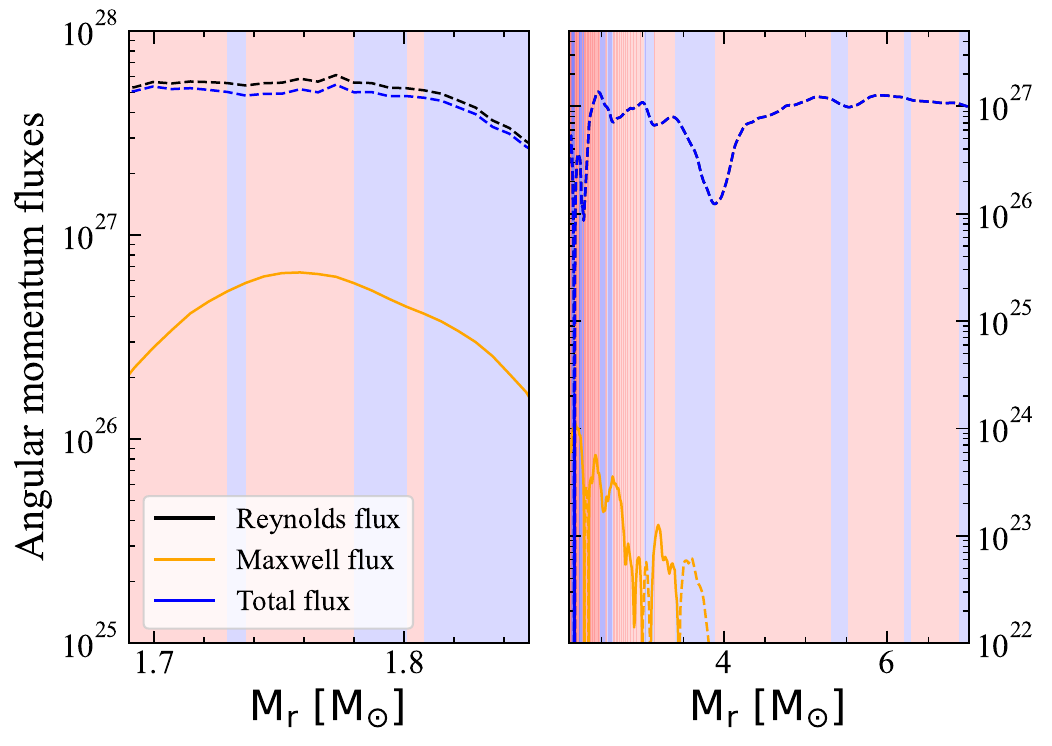}
    \caption{Angular-momentum fluxes in the  convective zones C1 (left) and C2 (right) of model \ad. The Reynolds contribution is the sum of Eqs.~\eqref{eqn:fluxes_adv}--\eqref{eqn:fluxes_aturb}, while the Maxwell contribution is given by Eq.~\eqref{eqn:fluxes_mag}. Solid lines represent positive fluxes and dashed lines negative fluxes. The background colouring is light red when $-\nabla_r F_{\rm tot}$ in Eq.\,\eqref{eq:AM_transport} is positive and blue when negative,  corresponding to local increases or decreases in specific AM, respectively. In the right panel the blue and black dashed lines (i.e., the Reynolds and total fluxes) overlap.}
    \label{fig:fluxes_ang}
\end{figure}

Figure~\ref{fig:fluxes_ang} shows the Reynolds flux, the Maxwell flux, and their sum for regions C1 and C2 of model \ad. The fluxes are averaged over a time window of one convective turnover once the turbulent flow has reached a quasi-stationary state.
%MAA: no need of footnote to specify the way in which the temporal average is done.
%\footnote{This is done by taking a time $t_0$ when the turbulence is in a steady state and performing a temporal average over a window corresponding to the turnover time}.
%MAA: joined paragraphs and break into two the following one.
%
The Maxwell flux has the opposite sign to the Reynolds flux. However, in region C1 its amplitude is roughly an order of magnitude smaller than that of the Reynolds flux at the peak. Magnetic stresses therefore do not dominate the angular-momentum transport in this shell.
%MAA: new paragraph

The sign of $-\nabla_r F_{\rm tot}$, where $\nabla_r$ is the radial component of the divergence operator, determines whether a given layer gains or loses AM. In C1 the left side of the shell is predominantly red in Fig.~\ref{fig:fluxes_ang}, indicating a local increase in AM, while the upper part of the shell is predominantly blue, indicating a local decrease. This is precisely the behaviour seen in Fig.~\ref{fig:jr_omega}: the base of the convective shell gains AM and spins up, while the upper part of the shell loses it. The net effect is a profile closer to constant $j_r$ than to constant $\Omega$. 

The outer convective zone C2 shows a more fluctuating pattern. This is consistent with the fact that the region has not yet reached a quasi-stationary state. A longer simulation is required to determine whether C2 ultimately approaches the same constant-$j_r$ configuration as C1, or whether magnetic stresses become more important once the turbulent state is fully established.

For progenitor \ad, the Maxwell flux in C1 is too weak to counteract the hydrodynamic tendency toward a $j_r$-constant profile. Since the Maxwell stress scales with the product $B_rB_\phi$, an increase of both magnetic-field components by a factor of order a few would make the magnetic flux comparable to the Reynolds flux. In that case, the sign of the total flux divergence could reverse, allowing the shell to evolve back toward an $\Omega$-constant profile.

The direction of AM transport in this convective region is finely balanced and only a small amount of amplification would reverse it. The magnetic field strengths in region C1 of \ad reach saturation strength, i.e. they will not grow further, so longer simulations are unlikely to change the distribution of AM we observe in the final model. Consequently, angular-momentum profiles of the stellar interior inferred directly from one-dimensional SE models may differ significantly from those obtained after a multidimensional MHD pre-collapse evolution.

\subsection{Magnetic-field amplification}
\label{sec:magnetic_amplification}

The magnetic field is amplified in all regions of both progenitors, but the dominant amplification mechanism differs between radiative and convective layers. In the radiative regions, the growth is modest and mainly associated with stellar contraction and the winding of the poloidal field into toroidal field by differential rotation. In convective regions, by contrast, turbulent motions transport magnetic flux from neighbouring magnetised layers and amplify both the poloidal and toroidal components. A cautionary note is warranted here. Owing to the limited resolution of 3D models, they may not fully capture the growth of the TS dynamo or the MRI, see Appendix~\ref{sec:MRI_resol} for further discussion. It is therefore quite possible that the magnetic-field amplification in the radiative regions of our models underestimates the true field growth in those layers. The field amplification reported here should therefore be interpreted as the result of the resolved multidimensional MHD evolution, supplemented by the numerical dissipation of the scheme, rather than as a complete description of all possible small-scale magnetic-instability growth in the stellar interior.  

In the SE models, convective regions are assumed not to retain coherent magnetic fields after many turnover times, and therefore no magnetic field is prescribed in those regions (Fig.~\ref{fig:B_and_O}). Consequently, after mapping an SE model into a multidimensional MHD calculation, turbulent layers either remain initially unmagnetised, if they coincide with convective shells in the SE model, or contain only the field inherited from layers that are classified as radiative in the SE data but later become turbulent in the multidimensional evolution.

In model \ag, region C1 develops turbulence in the multidimensional calculation even though it is not formally flagged as convective at the mapping time of the SE model, as discussed in \citetalias{Griffiths2026PaperI}. In that case, the SE model provides an initial magnetic field in region C1. We tested 2D models with and without this starting field in region C1 and found that the final field strength is controlled by the subsequent turbulent amplification rather than by the precise initial seed. In a statistical sense, quantified by the decomposition of the field in spherical harmonics (see Sect.~\ref{sec:B-topology}), the field geometry was also independent of the initial choice as it was entirely driven by the turbulent motions occurring in the convective region. A similar situation occurs in C1 of \ad, where non-zero SE magnetic-field estimates are present despite the layer being flagged as convective in SE.\footnote{The origin of this inconsistency in the \MESA model is unclear. Since the final field in the convective layers does not depend sensitively on the seed field, we initialise this layer with the SE field provided by the model.}
%MAA: said above
These tests indicate that, once turbulence develops, the final magnetic field in the convective shells is largely insensitive to the detailed seed field, provided that magnetic flux is available from the neighbouring radiative regions.

\begin{figure*}[t!]
    \centering
    \includegraphics[width=0.8\textwidth]{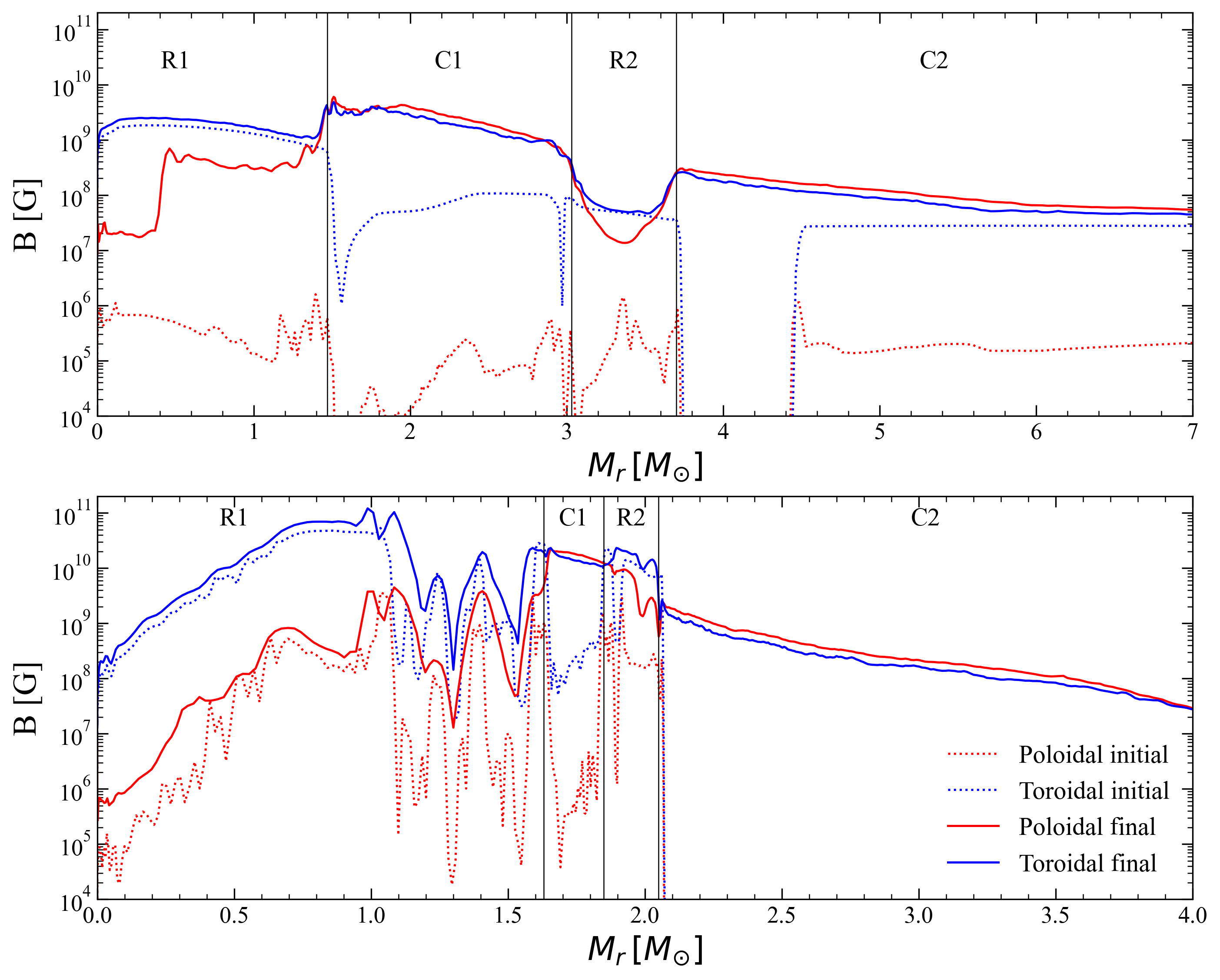}
    \caption{Angular-averaged poloidal and toroidal magnetic fields at start (dotted lines) and end (solid lines) of multi-D evolution for models \ag (top) and \ad (bottom). The labels R and C denote radiative and convective regions, respectively, following the notation of \citetalias{Griffiths2026PaperI}.}
    \label{fig:magnetic_field_start_end}
\end{figure*}

%MAA: full stop added
Figure~\ref{fig:magnetic_field_start_end} compares the angular-averaged poloidal and toroidal field strength at the beginning and end of the multi-D evolution. We point out the existence of two regions with no initial magnetic field as they are convective in the corresponding SE models.  They are both located in the C2 regions, namely, from $\simeq 2\,M_{\odot}$ outward in model \ad and between 3.7$\,M_{\odot}$ and 4.2$\,M_{\odot}$ in model \ag.
During the 3D evolution magnetic flux is advected into these regions from the adjacent radiative layers and subsequently amplified by turbulent motions. As a result, regions that are magnetically disconnected in the 1D initial estimate become magnetically linked in the multidimensional models.

%The situation for \ad is slightly different. Here, the C1 region is identified as convective, yet, the SE model\footnote{Recall here the code employed for this model is MESA.} still provides diffusion coefficients for the TS dynamo and the corresponding magnetic-field estimates. To avoid introducing an arbitrary initial field in that convective layer, we adopt the SE magnetic-field estimates, even though these correspond to saturation values appropriate for radiative layers. However, as in \ag, the subsequent turbulent amplification renders the initial seed field largely irrelevant for the final magnetic strength. 

In the inner convective regions the  initial poloidal and toroidal components increase by up to five orders of magnitude in some cases. By the end of the simulations, these regions have reached a quasi-saturated state in which the toroidal and poloidal components attain comparable strengths. The outer convective regions are not as close to reaching their asymptotic states, especially the upper part of C2 in \ad, where the turbulent flow does not reach the entire oxygen shell before collapse.\footnote{See \citetalias{Griffiths2026PaperI} for discussions on the development of convection in that layer.}  Nevertheless, the base of this shell reaches field strengths close to $10^9$\,G, even though the SE prescription initially assigns no magnetic field there.

Although some amplification of the poloidal field occurs in the radiative core and surrounding radiative shells, it generally remains weaker than the toroidal field. In the cores of both models, we observed only a modest increase in toroidal field strength, while the poloidal component is typically one to two orders of magnitude smaller.
%MAA: full stop

The presence of magnetic fields in convective regions stands in stark contrast to the usual 1D SE treatment, in which magnetic fields are not prescribed in these layers. In classical flux-expulsion arguments, sustained convection can remove coherent magnetic flux from convective cells. Whether such expulsion operates efficiently in the present global 3D progenitor models is less clear. The effective magnetic Reynolds number in our simulations\footnote{This is estimated using Fig.16 of \cite{Rembiasz_2017ApJS..230...18} to obtain $\eta_{\rm mag}\approx10^8$ and then $R_m \sim \frac{V_{\rm conv}H_p}{\eta_{\rm mag}}$.} is of order $10^7$. Since the flux-expulsion timescale increases with $R_m$ \citep{1966_Weiss}, the limited duration simulated prior to collapse  may be insufficient for complete expulsion to occur. Moreover, the 3D turbulent flow continuously folds, stretches, and reconnects field lines through the effective numerical resistivity, providing additional pathways for maintaining magnetic flux in the convective layers.

A useful diagnostic of the magnetic-field structure is the winding length scale, defined as
\begin{equation}
    \lambda = \frac{|\mathbf{B}|}{|\nabla \times \mathbf{B} |}.
\end{equation}

 In the inner convective shell C1 of \ad, we find $\lambda/\Delta r \sim 1.5$, where $\Delta r$ is the local grid size. Thus, the dominant winding scale remains resolved, although only marginally. As we do not explicitly include magnetic resistivity, dissipation occurs through the effective numerical resistivity of the scheme. 
 %\delt{The field lines would need to be wound down to much smaller scales (namely, close to the numerical cells size) before numerical dissipation could remove it efficiently. This supports the interpretation that the magnetic fields found in the convective regions are not a short-lived numerical artefact, although}\maaC{the fact that $\lambda\sim 1$ implies that we are marginally above the "acceptable limit". Thus, this cannot support your point here. } 
 %Ok I thought above 1 was sufficient.
 Higher-resolution and longer-duration simulations would therefore be required to determine how robustly the small-scale magnetic structure survives and whether complete flux expulsion could occur on longer timescales.

%Both components of the magnetic field saturate at about $10^9$ G for \ag and $4\times10^9$G for \ad. The field then roughly follows a $r^{-2}$ slope in the convective region, except the base of C1 for \ag where the magnetic field strength is flat. This slope tracks with the value of shear in the region where $q = \frac{\rm dln \Omega}{\rm dln r} \sim -2$ due to $j$-constant rotation in the shell. 

\citet{Varma_Mueller_2023} note that the magnitude of their saturated convective field is consistent, to within a factor of a few, with the saturation field strength expected from the MRI \citep{2003_Akiyama},
\begin{equation}
\label{eqn:B_sat_MRI}
    B^2_{\rm sat}\sim  \rho r^2 \Omega^2 \left| \frac{d\ln \Omega}{d \ln r}\right|.
\end{equation}
This estimate also provides a good description of the saturation strength in the convective regions in our models. For instance, the saturation value predicted by Eq.~\ref{eqn:B_sat_MRI} in C1 of \ad is $1.9 \times 10^{10}$\,G, whereas the average total field at the end of the simulation is $1.8\times10^{10}$\,G. However, as also pointed out by \cite{Varma_Mueller_2023}, this agreement should not be taken as evidence that the MRI is operating in the convective shell. Indeed, the MRI instability criterion,\footnote{See Eq.~6 of \cite{Griffiths_2022}} is generally not satisfied in the convective regions of the SE models. Rather, Eq.~\ref{eqn:B_sat_MRI} provides a useful order-of-magnitude estimate for the saturation of a field amplified by radial differential rotation and turbulent shear (i.e., by magneto-shear). 
% MAA: full stop

\def\arraystretch{1.1}
\begin{table*}
\centering
\caption{Magnetic-field amplification measured using Lagrangian tracer particles in the 3D simulations. For each region, we report tracer-averaged quantities relevant to the magnetic-field evolution. A quantity denoted by $\overline{X}$ corresponds to an average over tracers, while $\langle \cdot \rangle$ indicates an additional temporal average taken over a time interval during which the turbulent behaviour is approximately stationary.}
\begin{threeparttable}
\begin{tabular}{ c c c c c c c c c c }
\hline 
\hline
Models & \multicolumn{4}{c}{\ad} & & \multicolumn{4}{c}{\ag} \\\cline{2-5}\cline{7-10}
& R1 & C1 & R2 & C2 & & R1 & C1 & R2 & C2 \\
$r_{\rm min} [10^8 \rm cm]$ & 0.2 & 2.8 & 4.1 & 5.5 & & 0.5 & 2.5 & 9 & 12 \\
$r_{\rm max} [10^8 \rm cm]$ & 1.5 & 3.6 & 4.9 & 35 & & 3.9 & 5.5 & 11 & 20 \\
$B_{\rm max,ini} [10^{10} \rm G]$ & 15 & 5.3 & 5.6 & 0.14 & & 4.9 & 6.1 & 1.3 & 0.077 \\
$B_{\rm max,fin} [10^{10} \rm G]$ & 35 & 15 & 9.1 & 1.6 & & 7.7 & 7.5 & 2.1 & 0.32 \\
\hline
$\overline{r}_{\rm ini} / \overline{r}_{\rm fin}$ & 1.86 & 1.41 & 1.37 & 0.58 & & 1.94 & 1.30 & 1.19 & 0.69 \\
$\overline{B}_{\rm pol,fin} / \overline{B}_{\rm pol,ini}$ & 14.1 & 39 & 26.5 & 1200 & & 4.99 & 2.99 & 3.02 & 4.66 \\
$\overline{B}_{\rm tor,fin} / \overline{B}_{\rm tor,ini}$ & 3.75 & 21 & 2.45 & 18 & & 6.31 & 12.2 & 4.38 & 10.3 \\
\hline 
$\langle \overline{\rm Ro}_{\rm f}\rangle$ &  NA & 8 & NA & 90 & & NA & 195 & NA & 300 \\
$\left\langle \overline{B}_{\rm tor}/\overline{B}_{\rm pol} \right\rangle$ & 47.6 & 0.877  & 4.55 & 0.741 & & 7.69 & 0.746 & 3.03 & 0.741 \\
\hline
\hline
\end{tabular}%
\end{threeparttable}
\label{tab:tracer_amplification}
\end{table*}

To quantify more precisely the magnetic-field amplification, we use $4\times 10^6$ Lagrangian tracer particles
%\footnote{For \ag, tracer particles were activated at 128\,s and therefore only record the final minute of \insr{the 3D} evolution. The amplification\delt{s} factors reported in Table~\ref{tab:tracer_amplification} for this model thus correspond to the interval from 128\,s to collapse, not to the full simulation.} 
that follow and record their instantaneous velocity, magnetic-field strength, and position at each time step.%
%MAA: shifted footnote to after the dot
\footnote{For \ag, tracer particles were activated at 128\,s and therefore only record the final minute of the 3D evolution. The amplification factors reported in Table~\ref{tab:tracer_amplification} for this model thus correspond to the interval from 128\,s to collapse, not to the full simulation.}
Table~\ref{tab:tracer_amplification} summarises the changes of the magnetic-field strength using tracer-averaged values within each region of interest.
%MAA: full stop

In the radiative core, the toroidal-field amplification is mostly explained by contraction. For example, in R1 of \ad the tracer-averaged radii, $\overline r$ decrease by a factor of $\overline{r}_{\rm ini}/\overline{r}_{\rm fin}\simeq 1.86$. The square of this factor is 3.46, close to the measured toroidal-field factor estimated by the ratio $\overline{B}_{\rm tor,fin}/\overline{B}_{\rm tor,ini}\simeq 3.75$ (Tab.~\ref{tab:tracer_amplification}), indicating that flux conservation accounts for most of the growth  in the radiative core. In contrast, the radiative shells R2 of both models exhibit toroidal-field growth larger than expected from contraction alone. This additional amplification reflects the exchange of magnetic flux with adjacent convective regions. An illustrative case is R2 of \ag (see Fig.~\ref{fig:magnetic_field_start_end}), where the field at its boundaries grows until it approaches the strengths reached in the neighbouring convective zones, see Appendix~\ref{sec:magbnd_evol} for more details. 

The ratio between the toroidal and poloidal magnetic field strengths also differs between radiative and convective layers. Radiative regions remain toroidally dominated ($\langle\overline{B}_{\rm tor}/\overline{B}_{\rm pol} \rangle>1$), whereas convective regions evolve toward approximate equipartition between the two components. Region C1 of \ad yields the closest ratio to 1, with the other convective regions seemingly converging towards $\sim 3/4$. This behaviour reflects the relatively rapid rotation in C1 of \ad, which enhances the winding of turbulent poloidal field into toroidal field. This can be characterised by the fluid Rossby number,
\begin{equation}
\label{eqn:Rof}
    \rm Ro_{\rm f} = \frac{|\omega|}{2\Omega},
\end{equation}
where $\omega =\nabla \times v$ is the local vorticity.\footnote{This definition follows \cite{Noraz_Brun_Strugarek_2024} and generally leads to higher values than the definitions used in \cite{Varma_Mueller_2023} and \cite{Shimada_McNeill_Varma_Maeda_Yokoyama_Muller_2026} for example.} 
Therefore, for sufficiently high Rossby numbers, $\rm Ro_{\rm f} \ga   90$,  $\langle\overline{B}_{\rm tor}/\overline{B}_{\rm pol} \rangle$  converges towards $\sim 3/4$, i.e., to a slightly poloidally dominated configuration. Only in region C1 of \ad, with a much lower Rossby number ($\rm Ro_{\rm f}\simeq 8$), does the ratio converge to a higher value. From the limited set of convective regions studied here we find a weak trend $\langle \overline{B}_{\rm tor}/\overline{B}_{\rm pol} \rangle\propto \rm Ro_{\rm f}^{-0.09}$. Given the small sample size, the limited numerical resolution, and the relatively large Rossby numbers of most regions, this scaling should be regarded as indicative rather than definitive. It nevertheless suggests that faster rotation favours a slightly more toroidally dominated saturated field, while slowly rotating convective shells approach a nearly balanced toroidal--poloidal configuration.

\subsection{Magnetic field topology}
\label{sec:B-topology}

\begin{figure}[t!]
    \centering
    \includegraphics[width=\columnwidth]{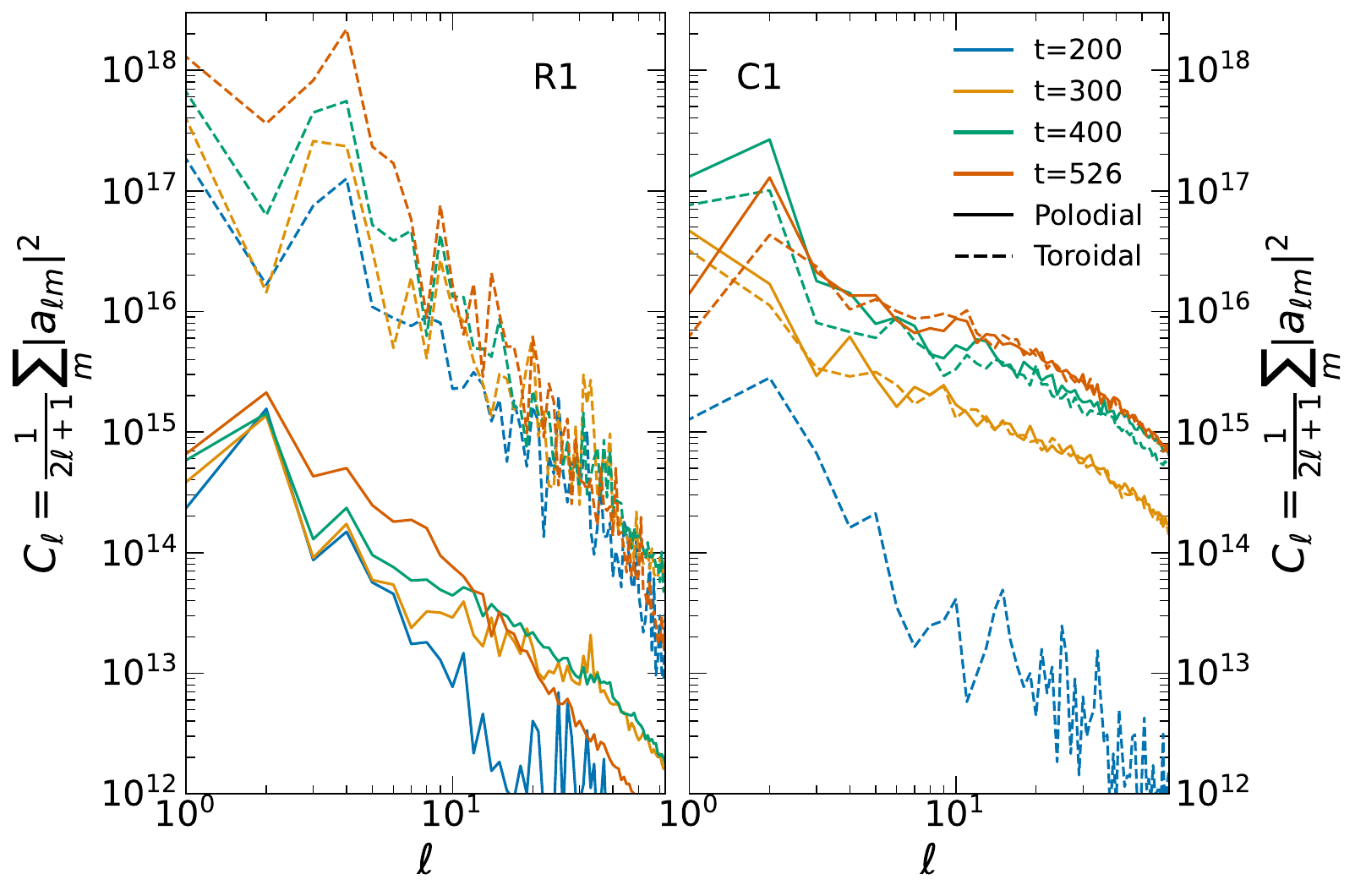}
    \caption{Spherical-harmonic decomposition of the poloidal (solid lines) and toroidal (dashed lines) magnetic field components for model \ad in the radiative core R1 (left) and in the convective zone C1 (right).}
    \label{fig:spectral_ad13}
\end{figure}

By the time the progenitors reach the pre-SN link, their deep interiors are strongly magnetised. The resulting topology is, however, not uniform throughout the star. Radiative regions tend to preserve a large-scale, predominantly toroidal field, whereas convective regions develop a more complex multiscale structure with substantial power at intermediate and high spherical-harmonic degrees.

To quantify this difference, we analyse the spherical-harmonic decomposition of the toroidal and poloidal magnetic-field components. For a scalar field $f(r,\theta,\phi)$, representing either the toroidal component, $B_\phi$, or the poloidal field strength, $B_{\rm pol}=(B_r^2+B_\theta^2)^{1/2}$, we define 
\begin{equation}
\label{eq:a_lm}
    a_{\ell m} ( r )  = \int_{\Omega} f(r,\theta, \phi)\, Y_{\ell m}^*(\theta, \phi) \, d\Omega,
\end{equation}
where $Y_{\ell m}$ are the complex spherical harmonics and
\begin{equation}
\label{eqn:C_ell}
    C_{\ell} ( r ) = \frac{1}{2\ell +1 } \sum_{m=-\ell}^{m=\ell} |a_{\ell,m}|^2 (r).
\end{equation}
To reduce stochastic fluctuations and to track the same physical layers as the star contracts, the spectra are  averaged over a small mass interval within each region of interest, rather than over a fixed radial interval. 

Figure~\ref{fig:spectral_ad13} shows the resulting spectra for model \ad in the radiative core R1 and in the inner convective shell C1.
In the radiative core, the toroidal component dominates over the poloidal one. The poloidal-field spectrum peaks at $\ell=2$, whereas the toroidal-field spectrum peaks over the range $\ell\simeq1$--$4$.
%\maaC{Adam, note the change!}
%Yes 4 does sometimes overtake 1 so stating both is ok
This large-scale structure remains broadly unchanged from the beginning of the 3D run--after 200\,s of 2D evolution--to collapse. The main evolution observed in the spectra is in an overall increase in magnetic power, primarily at low multipoles, together with a modest transfer of power toward higher~$\ell$.
%At intermediate times, the radiative layer exhibits a flattened plateau at moderate $\ell$ values, followed by a power-law decline for $\ell \gtrsim 30$. By the end of the evolution, the spectrum develops a remarkably smooth power-law dependence across almost the entire range of multipoles, extending down to $\ell \simeq 2$, with a slope very similar to that observed at high $\ell$ throughout the simulation. 

The convective region C1 behaves very differently. The initial geometry, inherited from the axisymmetric setup after the 2D relaxation phase, is rapidly reshaped once the fully 3D turbulent evolution begins.
%\footnote{Due to the impossibility of winding the magnetic field in 2D evolution the turbulence amplification mechanism does not occur.}
Between $t=200$\,s and $t=300$\,s, the poloidal and toroidal components both undergo substantial amplification and reach comparable power. The resulting spectrum is flatter than in the radiative core, with a peak at $\ell=2$ for both components, and contains significant power over a wide range of $\ell$. This reflects the stretching, folding, and reorientation of field lines by turbulent convection. Similar behaviour is found in the other convective regions of our models. The contrast between R1 and C1 therefore shows that the final magnetic topology is not well represented by a single large-scale dipolar or toroidal component throughout the star.

%We have seen that during the 3D evolution the poloidal field is amplified in C1, which does not occur in a 2D run, and again here we see the two components share roughly the same power especially in high $\ell$'s. Both fields have a peak contribution around 2 where a quadrupolar geometry may be a good approximation. The power spectra in the turbulent region is nearly flat and maintains a larger amount of power in the high $\ell$'s showing that the field has a significant small scale structure when in turbulent flows. 
%At the time of collapse in the poloidal field in R1 decays faster in the low $\ell$s with in an increase in power at the peak due to stretching of the field. The geometry of region C1 stays the same at collapse maintaing a lot of power in high $\ell$'s. We therefore see a clear difference in the geometry of the fields in these regions  with a larger scale field for radiative regions and a smaller scale one in convective.
%Although in region C1 of \ag some decay of the small scale field begins as convection is slowly shutting down by the time of collapse as it is no longer fuelled by strong nuclear burning. Thus without the active contribution of turbulence we see that the small scale field will decay back into the shape of the field observed in radiative zones. 
\begin{figure}[t!]
    \centering
    \includegraphics[width=\columnwidth]{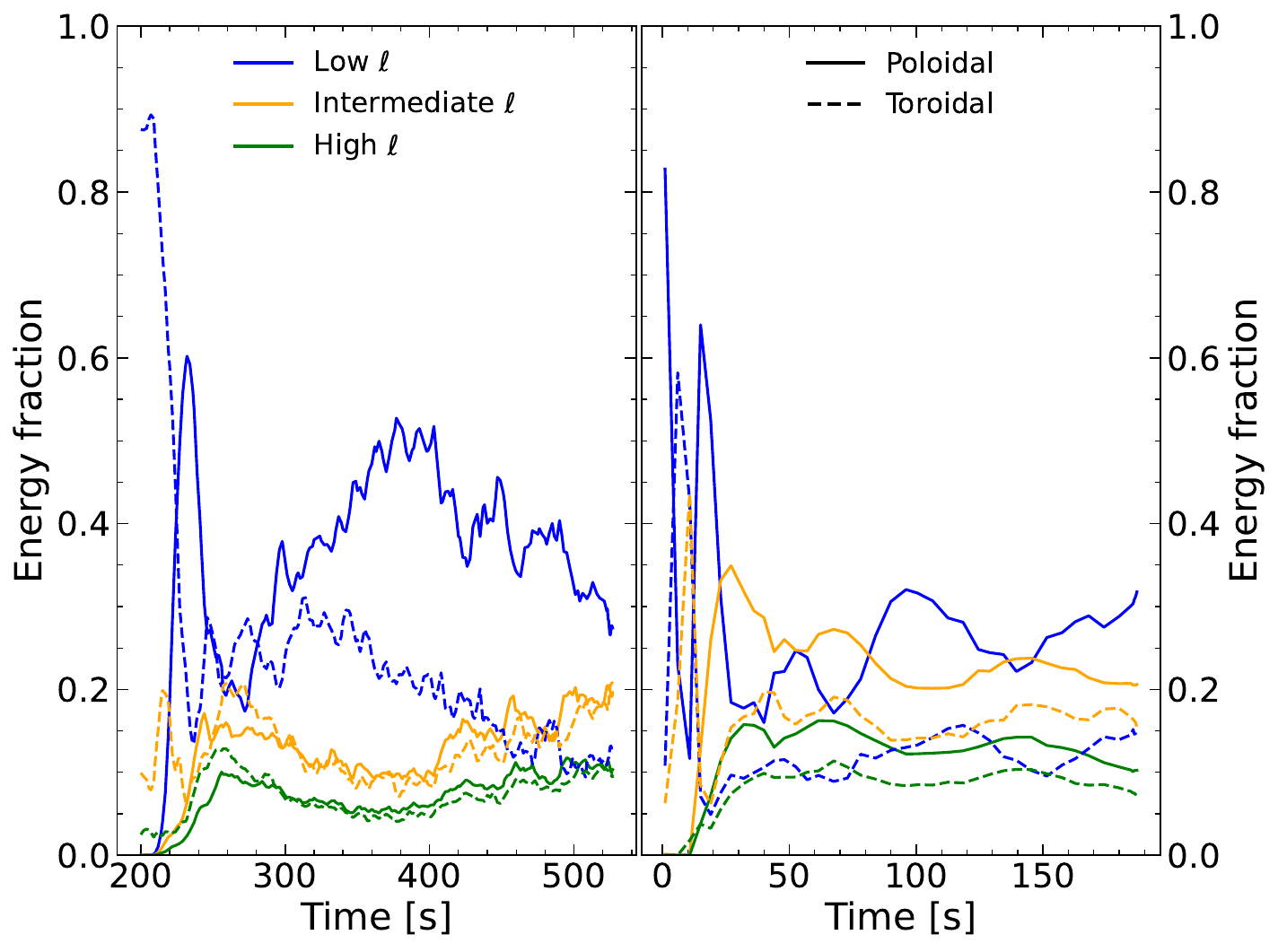}
    \caption{Evolution of the normalised magnetic-energy distribution in the C1 zones of model \ad (left) and \ag (right). Poloidal contributions are shown with solid lines, and toroidal contributions with dashed lines, each normalised by the  total magnetic energy. The energy is decomposed into three  ranges of spherical harmonic degree: a low-$\ell$ range  ($1\leq\ell\leq5$), an intermediate-$\ell$ range ($5<\ell\leq25$), and a high-$\ell$ range ($25<\ell$).}
    \label{fig:energy_sphharm}
\end{figure}

To follow the redistribution of magnetic power in time, we group the magnetic energy into three spherical-harmonic ranges: low $\ell$, with $1\leq\ell\leq5$; intermediate $\ell$, with $5<\ell\leq25$; and high $\ell$, with $25<\ell$.
%MAA: footnote integrated in the text and rewritten to soften the arbitrariness of the choice.
%\footnote{
These ranges are not meant to define physically sharp regimes, but they provide a useful diagnostic of the transfer of magnetic energy from large-scale to smaller-scale structures.
%}
%MAA: new paragraph

Figure~\ref{fig:energy_sphharm} shows this decomposition for the C1 regions of both 3D models. In both progenitors, most of the magnetic power initially resides at low~$\ell$, reflecting  the large-scale field imposed in the initial conditions and preserved during the preceding 2D relaxation phase. Once the 3D turbulent evolution begins,
%MAA: footnote integrated in the text
%\footnote{After 200\,s of 2D evolution, the toroidal component of model \ad still retains most of its power at low~$\ell$,  again highlighting how the 2D phase preserves the dipolar structure of the initial field--even in convective regions.} 
a significant portion of the magnetic energy rapidly shifts to the intermediate- and high-$\ell$ ranges. 
In model \ad, the high-$\ell$ contribution eventually dominates, with the poloidal and toroidal components each contributing close to 30\% of the total magnetic energy. In model \ag, the low-$\ell$ modes remain the largest single contribution, but the power in these modes is strongly reduced compared to its initial value, and the intermediate- and high-$\ell$ ranges acquire a significant fraction of the total power.

%MAA: another paragraph for the behavior in radiative regions
This redistribution is largely confined to convective layers. In radiative regions, not shown in Fig.~\ref{fig:energy_sphharm}, over 90\% of the magnetic energy remains in low-$\ell$ modes throughout the evolution. The final progenitors therefore contain two distinct types of magnetic topology: large-scale, predominantly toroidal fields in radiative regions, and multi-scale, approximately toroidal--poloidal fields in convective shells. A complementary visualisation of this structure is shown in Appendix~\ref{sec:3D_vis}.
%We clearly see the multi-scalar geometry of the magnetic field in convective regions which is well contrasted with the radiative zones. 

%MAA: full stop

%
%\maaC{Move it to the discussion.}
%\insr{If this structure survives during infall, it may affect the magnetic topology accreted onto the proto-neutron star and the large-scale field available to launch or collimate a magnetorotational outflow. Assessing this requires collapse simulations initialised from the present 3D progenitors.}
%MAA: full stop

\section{Discussion}
\label{sec:discussion}

%\maaC{The structure of the discussion is already very good, but there is some speculative wording that should be made more careful.}

The models presented here show that multidimensional MHD evolution can substantially modify the rotational and magnetic structure inherited from one-dimensional SE calculations during the final minutes before core collapse. Two results are particularly relevant. First, angular-momentum transport in convective regions does not necessarily preserve the nearly rigid rotation predicted by one-dimensional SE models; in our simulations, Reynolds stresses drive the flow toward an approximately constant specific-angular-momentum profile, while Maxwell stresses oppose this tendency but do not dominate it. Second, convective shells that are initially unmagnetised in the SE prescription become magnetised through flux transport from neighbouring radiative layers and subsequent turbulent amplification.

These results do not yet define a universal prescription. They are based on two progenitors and on a limited number of convective shells, with the outer shells less well sampled than the inner ones. Nevertheless, they provide useful guidance for constructing more realistic core-collapse initial conditions and for improving the treatment of magneto-convective angular-momentum transport in 1D SE.

\subsection{Implications for core-collapse initial conditions}
%MAA: modified section title
%\subsection{Initialising core-collapse models}

The 3D progenitors constructed here are intended for future collapse and explosion calculations. However, producing fully multidimensional pre-collapse models for a broad grid of stellar masses, metallicities, rotation rates, and magnetic-transport prescriptions is not computationally feasible. A practical alternative is to use multidimensional calculations such as these to guide the construction of improved initial conditions from 1D SE models. 

The first implication concerns the angular-velocity distribution. Stellar-evolution models usually assume shellular rotation, $\Omega=\Omega(r)$. Our two-dimensional test calculation, however, shows that the rotation profile can reorganise toward a more cylindrical structure during the final pre-collapse evolution. 
The $\phi$-averaged 3D profile shown in Fig.~\ref{fig:omega_spherical_vs_2d} further shows that this cylindrical-like organisation is not erased by the subsequent fully 3D turbulent evolution.
Moreover, within convective regions the 3D simulations tend to develop approximately constant-$j$ profiles rather than the nearly constant-$\Omega$ profiles predicted by SE. Mapping a 1D progenitor directly into a core-collapse calculation with a purely shellular rotation law may therefore misrepresent the angular-momentum distribution in the material that later accretes onto the proto-neutron star.

Rotation laws approaching constant specific angular momentum at large cylindrical radius have been used previously in idealised magnetorotational core-collapse simulations. A standard example is the classical ``$j$-constant'' law, $\Omega(\varpi) =  \Omega_0/(1+(\varpi/A)^2)$ \citep[where $A$ is a constant;][]{Eriguchi_Mueller_1985A&A...146..260, Obergaulinger_Aloy_Mueller_2006A&A...450.1107, Obergaulinger_Aloy_Dimmelmeier_Mueller_2006A&A...457..209}. 
More recent magnetorotational core-collapse simulations often impose other parametric differential-rotation profiles \citep[e.g.][]{Takiwaki_Kotake_Nagataki_Sato_2004,Kuroda_Arcones_Takiwaki_Kotake_2020,Shankar_Mosta_Haas_Schnetter_2025}. 

The AM distribution in the pre-SN model is especially relevant for magnetorotational explosions. It controls the centrifugal support of the collapsing core, the winding of magnetic field lines, the rotational energy available to the proto-neutron star, and the conditions for jet formation. A cylindrical or constant-$j$ redistribution changes where angular momentum is concentrated relative to the rotation axis and to the mass shells that accrete earliest. The quantitative impact of this redistribution must be assessed with collapse simulations, but the present results suggest that the rotational mapping from SE models should not be treated as a purely technical choice.

The second implication concerns the magnetic field. In standard SE models, magnetic-field estimates are typically available only in radiative regions, while convective shells are often assigned no field. Our simulations show that this can produce an artificially disconnected initial magnetic structure. In the multidimensional evolution, magnetic flux is transported into convective shells and amplified there until the toroidal and poloidal components reach comparable strengths. Thus, collapse calculations initialised directly from SE models should not necessarily leave convective regions unmagnetised.

A possible practical approach is the following. 
In radiative regions one may use the saturation-field estimates provided by the SE magnetic-transport prescription, while keeping in mind their order-of-magnitude nature. In convective regions, one may initialise a non-zero magnetic field using a shear-based saturation estimate such as Eq.~\ref{eqn:B_sat_MRI} as an order-of-magnitude estimate of the field strength reached by turbulent shear amplification. The relative toroidal and poloidal components could then be chosen close to equipartition, as suggested by the convective regions in our models. For slowly rotating convective shells, our simulations indicate $\langle \overline{B}_{\rm tor}/\overline{B}_{\rm pol} \rangle \simeq 0.7-0.8$,  while the more rapidly rotating C1 shell of \ad approaches a ratio closer to unity.\footnote{The weak Rossby-number trend found in Section~\ref{sec:magnetic_amplification} should be regarded as indicative rather than as a calibrated scaling.} To test the robustness of this scaling, it will be necessary to examine additional progenitors with burning shells spanning a wider range of Rossby numbers, including values closer to, or even below unity. Achieving such conditions may be difficult during late burning stages. In SE models evolved with magnetic-field-driven AM transport, the core is significantly spun down compared with hydrodynamic models, while convection during silicon and oxygen shell burning remains very rapid. One possible avenue to reach lower Rossby numbers is through binary mergers, which can produce rapidly rotating cores even when magnetic AM transport is active. Such systems might yield turbulent shells with lower Rossby numbers and, perhaps, correspondingly weaker poloidal fields relative to toroidal fields. 

The topology of the reconstructed field is also expected to be important for the post-bounce dynamics. Radiative regions in our models preserve most of their magnetic energy at low spherical-harmonic degree, whereas convective regions develop substantial power at intermediate and high~$\ell$. 
The present spectra suggest that low-order poloidal components, in particular $\ell=2$, may be important. However, higher-resolution simulations that better capture small-scale instability growth may modify this conclusion.
In the convective regions, the field contains much more power on small scales. Therefore, the final power spectra of our 3D models may provide a useful guide for initialising generic fields in the convective regions of new progenitors.

%MAA: To introduce the comparison with Gottlieb et al., let us recall a basic shell-merger result.
Before discussing magnetic-flux survival during early accretion, it is useful to recall a structural result from \citetalias{Griffiths2026PaperI}. In the original 1D model of \ad, the Si- and O-burning shells merge before collapse, whereas the 3D MHD evolution preserves a distinct Si-burning C1 shell separated from the O-burning C2 shell by the radiative layer R2. The discussion of magnetic-flux survival below therefore refers to the multidimensional pre-collapse structure, not simply to the original 1D SE profile.

To assess the longevity of the magnetic field in convective regions, the comparison with \citet{Gottlieb_Renzo_Metzger_Goldberg_Cantiello_2024}
is particularly relevant for model \ad, which is both compact and rapidly rotating. In this model, the radiative iron core is expected to form the initial proto-neutron star, while the first layers to accrete after bounce belong to the silicon-burning shell C1, the intervening radiative layer R2, and then the inner part of the oxygen-burning shell C2. This ordering matters
for the interpretation of magnetic-flux survival. The argument of
\citet{Gottlieb_Renzo_Metzger_Goldberg_Cantiello_2024} is most directly applicable to magnetic fields generated locally by small-scale AM-transport instabilities in radiative layers, where short
coherence lengths, incoherent polarity, and reconnection may strongly reduce the net large-scale flux delivered to the compact object. In \ad, however, the radiative layer R2 between C1 and C2 occupies only a relatively narrow mass interval, whereas most of the material accreted once the proto-neutron star reaches $M\gtrsim1.6\,M_\odot$ comes from convective layers.

Our 3D pre-collapse evolution suggests that these convective layers should not be treated in the same way as radiative layers whose magnetic fields are generated locally by small-scale AM-transport instabilities. In C1 and in the inner part of C2, magnetic flux is transported from neighbouring radiative regions and then amplified and maintained by the resolved turbulent flow. Once the turbulent state is established, the magnetic-energy spectrum and
the relative distribution of power over spherical-harmonic degree evolve only moderately before collapse. Thus, on the resolved scales of our simulations, the magnetic structure in the convective layers is not erased before the pre-SN link. Since these layers are expected to dominate the early post-bounce accretion after the initial core has formed the proto-neutron star, their
magnetic field may provide a more robust contribution to the magnetic flux accreted by the compact object than would be inferred from a purely 1D radiative-instability picture.

This suggests that the pre-collapse magnetic structure relevant for early accretion may be less vulnerable to the flux-cancellation argument than a purely radiative, small-coherence-length field. This does not imply that the net large-scale flux reaching the proto-neutron star is guaranteed to be conserved: the convective fields are multiscale and locally incoherent, and the accreted flux will depend on how this structure is compressed, advected, reconnected, and wound during infall and post-bounce evolution. In compact rapidly rotating progenitors like \ad, a significant fraction of the early accreted
mass is magnetised by convective turbulent amplification before collapse, and
this field survives until the pre-SN link in the present 3D MHD simulations. Whether this flux contributes coherently to the proto-neutron star field can only be assessed with collapse simulations initialised from the present 3D models.

%\textcolor{blue}{AG: building a magnetic field }

%The reconstruction of a realistic magnetic field for core-collapse progenitors must therefore account for the small-scale topology generated by turbulent amplification in convective regions. A multipolar reconstruction is required, rather than the commonly assumed dipolar geometry. Likewise, the more stable field configuration in radiative zones at collapse suggests that a quadrupolar reconstruction could be more appropriate for the poloidal field in those layers.

%Based on the amplification in our convective regions,  we suggest a possible approach to build initial magnetic fields for convective zones when mapping 1D SE models to core-collapse simulations. As in \cite{Varma_Mueller_2023}, we find that the field saturates at values consistent with the MRI saturation estimate of \cite{2003_Akiyama}. This relation may therefore provide a reasonable estimate of the toroidal field strength in a convective region. Given the Rossby number inferred from SE data,  the poloidal component could then be scaled according to the empirical relation we identify here. Although admittedly uncertain, such an approach would provide at least a different scaling between fast and slow rotating regions. This, in turn, would allow core-collapse simulations to be initialised with a more realistic magnetic field in convective regions--certainly more likely than assuming no field at all. 

 %\subsection{Magneto-convection for SE modelling}

%MAA: an alternative title in veo 
\subsection{Implications for stellar-evolution modelling}
\label{sec:discussion_SE}

%Figure moved to the appendix

The presence of magnetic fields in convective regions also has implications for 1D SE calculations. In SE models, convection is usually assumed to redistribute angular momentum toward nearly uniform angular velocity. In our 3D simulations, hydrodynamic Reynolds stresses instead drive the convective flow toward an approximately constant-$j$ profile. Magnetic stresses act in the opposite direction and tend to restore a flatter $\Omega$ profile, but in our models they are not strong enough to overcome the Reynolds stresses.

Our results differ from the MHD model of \cite{Varma_Mueller_2023}, where Maxwell stresses developed in the convective regions become strong enough to reverse the hydrodynamic AM flux and restore a more rigidly rotating convective shell. The difference is not necessarily contradictory. The sign and magnitude of the net AM transport depend sensitively on the strength of the saturated magnetic field and on the correlation between the radial and toroidal field components, $B_rB_\phi$. In our C1 shell of \ad, increasing both magnetic-field components by a factor of only a few would make the Maxwell flux comparable to the Reynolds flux and could reverse the direction of the net transport.

%\maaC{Part of this text has been moved to Ap.\ref{sec:magbnd_evol}). It is of technical nature (on the influence of the boundary conditions "used by others".) So we could move there the figure 11 and, if you produce the other figure to show whether the evolution at the boundaries is simply given by flux conservation or something else. }
Several factors may therefore affect the final AM profile. One is the magnetic-field strength at the boundaries of the convective region.
%MAA: integrated in the text below
%\footnote{These simulations do not include the stellar core so boundary conditions are required and set far away from the convective/radiative boundary.} 
In simulations that excise part of the star, imposed magnetic boundary conditions may influence the field amplification inside the convective shell. \citet{Varma_Mueller_2023} set a fixed vertical-field of $10^7$\,G at their inner radial boundary. In our global models, the field is evolved self-consistently, and neighbouring radiative layers exchange magnetic flux with the turbulent shells. In Appendix~\ref{sec:magbnd_evol} we show the evolution of the poloidal and toroidal magnetic fields at selected boundaries of our convective regions.

A second factor is numerical resolution. The saturated field strength in turbulent MHD simulations can depend on the effective magnetic Reynolds number and on the numerical diffusivity of the scheme \citep[e.g.,][]{Varma_Muller_2026}. This is particularly relevant when the Maxwell stress is close to the threshold required to reverse the AM flux. Higher-resolution calculations, or calculations with controlled explicit diffusivities, would be needed to determine whether the saturation field in the present models is slightly underestimated. In this context, numerical diffusivity studies such as \citet{Rembiasz_2017ApJS..230...18} are directly relevant for assessing the reliability of magnetic-field amplification in global MHD simulations.
%

%\textcolor{blue}{AG: Also maybe can cite a Rembiasz or alike paper on saturation strengths? }
%MAA: done
%

The broader implication is that AM transport in convective shells should not be prescribed solely as a tendency towards rigid rotation. Instead, it may depend on the balance between hydrodynamic Reynolds stresses, which favour constant-$j$ profiles, and Maxwell stresses, which can favour flatter $\Omega$ profiles if the magnetic field becomes sufficiently strong.
Recent work by \cite{Shimada_McNeill_Varma_Maeda_Yokoyama_Muller_2026} points in this direction by relating Maxwell stresses in magneto-convective regions to local rotational and convective properties of the star. Our results support the need for such prescriptions, but also show that the field strength, topology, and boundary exchange of magnetic flux can be decisive. 

%\subsection{Impact of 3D on core-collapse/explosion \textcolor{blue}{AG: Better title required.}}
%MAA: here my alternative
\subsection{Possible impact on collapse and explosion}
\label{sec:discussion_collapse}

The impact of these multidimensional progenitors on the subsequent collapse and explosion remains to be quantified by dedicated core-collapse simulations. Nevertheless, the differences found here identify several channels through which the outcome may change relative to calculations initialised from the original one-dimensional SE models.

The first is the spin-up of the proto-neutron star. In model \ad, the inner convective shell C1 evolves toward a constant-$j$ profile, so the angular momentum contained in the lower part of the shell differs from the SE prediction. If a significant fraction of this shell accretes onto the proto-neutron star, the resulting spin-up can differ from that obtained from the 1D progenitor. This may affect the rotational energy available after bounce and the degree of differential rotation that can wind magnetic fields. 

The second is the magnetic topology accreted by the collapsing core. In the original SE magnetic-field estimate, the magnetised core and overlying magnetised layers are separated by convective regions where no field is prescribed. In the 3D models, these regions become magnetically connected. Convective shells also develop a substantial poloidal component and a multiscale, tangled topology. During collapse, such a field may be compressed, stretched, and wound into a different configuration from that obtained by collapsing a simple dipolar or toroidal field \citep{Aloy_2021, Obergaulinger_2021}.

The third is the magnetic field available for jet launching or collimation. Radiative regions retain a large-scale, predominantly toroidal structure, whereas convective regions contain a broader distribution of multipoles. Large-scale components may be more effective in organising an outflow \citep{Bugli_Guilet_Obergaulinger_2021}, while the role of small-scale components is not obvious (they may partly cancel or reconnect during infall). The balance between these effects will depend on how much of the convective-shell topology survives collapse and how it is processed by the proto-neutron star and post-shock flow.

%Finally, the field strengths reached within the inner iron core at collapse, boosted by both turbulent amplification and contraction, are sufficiently large to serve as viable seeds for magnetar-level fields, especially in the case of the faster-rotating \MESA progenitor. This statement should not be interpreted as a prediction of a magnetar-strength remnant by itself. The final field strength and topology will depend on collapse dynamics, post-bounce differential rotation, possible MRI growth, turbulent amplification, and magnetic reconnection. The present models instead provide a more physically motivated starting point for testing these processes in future multidimensional collapse simulations.

%%%%%%%%%%%%%%%%%%%%%%%%%%%%%%%%%%%%%%%%%%%%%%%%%%%%%%%%%%%%%%
\section{Conclusions}
\label{sec:conclusions}

We have analysed the rotational and magnetic properties of the first full 3D MHD pre-supernova progenitors evolved through the final minutes before core collapse. The global properties, turbulent shell structure, and nuclear-burning behaviour of these models were presented in \citetalias{Griffiths2026PaperI}; here we have focused on angular momentum redistribution, magnetic-field amplification, and magnetic topology.

Our main results can be summarised as follows. First, the AM distribution obtained from the multidimensional evolution can differ substantially from the 1D SE. In the inner convective shell of model \ad, Reynolds stresses drive the flow toward an approximately constant specific-angular-momentum profile, corresponding to an average rotation profile close to $\Omega\propto\varpi^{-2}$. Maxwell stresses oppose this tendency, but in our models they are not strong enough to restore the nearly rigid rotation usually assumed for convective regions in 1D SE calculations. The final pre-collapse rotation profile is therefore sensitive to the balance between hydrodynamic and magnetic stresses.
%MAA: moved here, as it was written after the amplification in convective regions.

Second, convective regions that are initially weakly magnetised, or even unmagnetised in the 1D SE prescription, acquire and amplify magnetic flux during the 3D evolution. Magnetic flux is transported from neighbouring radiative regions into the turbulent shells, where both the poloidal and toroidal components are amplified. In the convective regions, the saturated field approaches approximate toroidal--poloidal equipartition, with $B_{\rm tor}/B_{\rm pol}\simeq0.7$--$0.8$ in slowly rotating shells and a ratio closer to unity in the faster rotating C1 shell of \ad. 
%The weak trend with the fluid Rossby number found here should be regarded as indicative rather than as a calibrated scaling, given the limited number of convective regions sampled.
%

%MAA: original location of the paragraph on the amplification of the B-field in convective layers
%Angular momentum transport in convective regions does not follow the 1D predictions made by stellar evolution at these late times. The Reynold stresses induce a $j$-constant profile over convective regions whereas the stellar evolution models tend to predict $\Omega$-constant. The Maxwell stresses can counteract the Reynold stress to enforce rigid rotation once more but in our models the field strengths are not powerful enough. Our 3D models therefore arrive at the pre-SN link with a significantly different average rotation profile compared to what a 1D progenitor would predict.

Third, the magnetic topology differs markedly between radiative and convective layers. Radiative regions largely preserve a low-$\ell$, predominantly toroidal magnetic structure, whereas convective regions develop a multiscale field with substantial power at intermediate and high spherical-harmonic degrees. Thus, the pre-collapse field cannot be represented accurately by a single large-scale dipolar or toroidal component throughout the star. A more realistic construction of magnetic initial conditions for collapse simulations should account for both large-scale fields in radiative layers and turbulent, multiscale fields in convective shells.

Finally, our results suggest that multidimensional pre-collapse evolution can modify the magnetic connectivity of the progenitor. Regions that are magnetically disconnected in the 1D SE estimate become linked through flux transport and turbulent amplification. This may affect the magnetic flux and topology accreted by the proto-neutron star, the subsequent winding of field lines, and the conditions for magnetorotational outflows. Quantifying these effects requires collapse and post-bounce simulations initialised from the present 3D progenitors. Nevertheless, the models presented here already show that AM and magnetic-field structures inferred directly from 1D SE calculations may be incomplete, especially in the convective shells that accrete shortly after bounce.

\section{Data Availability}

The final 3D snapshots at the pre-SN link of each progenitor are stored on Zenodo \href{https://doi.org/10.5281/zenodo.19692957}{https://doi.org/10.5281/zenodo.19692957}. They will be made publicly available from 01/01/2027. For anticipated access please contact the corresponding author.

%%%%%%%%%%%%%%%%%%%%%%%%%%%%%%%%%%%%%%%%%%%%%%%%%%%%%%%%%%%%%%
\begin{acknowledgements}
We acknowledge support from grants PID2021-127495NB-I00 and PID2025-171322NB-C22, funded by MCIN/AEI/10.13039/501100011033 and by the European Union “NextGenerationEU". We also acknowledge support from the Astrophysics and High Energy Physics programme of the Generalitat Valenciana ASFAE/2022/026 funded by MCIN and the European Union NextGenerationEU (PRTR-C17.I1) as well as support from the Prometeo excellence programme grant CIPROM/2022/13 funded by the Generalitat Valenciana. 
\end{acknowledgements}

\bibliographystyle{aa}
\bibliography{biblio.bib}

\begin{appendix}

\section{Field strengths}
\label{sec:appen_field}
Magnetic instabilities are treated in stellar-evolution calculations through effective prescriptions. Their impact is on the diffusion of angular momentum and chemical species when appropriate. In \ag the description of the TS dynamo follows \cite{Eggenberger_2022} which represents a calibrated description of the original formalism proposed by \cite{Spruit_2002}. Model \ad describes the TS dynamo using the description of \cite{Heger_Woosley_Spruit_2005}. 

In the case of the TS dynamo as described by \cite{Eggenberger_2022} the saturation field expressions are\footnote{Note that in the Heaviside-Lorentz units that we use here there is no factor $\sqrt{4\pi}$.}
\begin{align}
\label{eq:B_phi_TS}
    b_{\rm tor} &= \sqrt{\rho} C_T |q| \Omega r \frac{\Omega}{N}, \\
\label{eq:B_r_TS}
    b_{\rm pol} &=\sqrt{\rho} C_T^2 q^2 \Omega r \left(\frac{\Omega}{N}\right)^3.
\end{align}
Here $q=\frac{\partial\ln \Omega}{\partial\ln r}$ is the shear of the rotating flow, $N$ the Brunt-Väisälä frequency and $C_T$ a calibration factor set to $216$. In the case of the \cite{Heger_Woosley_Spruit_2005} description no such calibration factor is applied and the expressions are the same as above but with $C_T = 1$.

\section{Reynolds and Favre decomposition}
\label{sec:appen_favre}
The Reynolds and Favre decompositions of a fluid quantity $X$ are respectively given by
\begin{equation}
    \hat{X} = \langle X\rangle = \frac{\int_{\omega} X d\omega}{\int_{\omega}d\omega} %MAA commented:\hspace{1cm} X' = X - \hat{X}
\end{equation}
and
\begin{equation}
    \widetilde{X}  = \frac{\int_{\omega} \rho X d\omega}{\int_{\omega}\rho d\omega} %MAA: commented\hspace{1cm} X'' = X - \widetilde{X}.
\end{equation}
where $d\omega=\sin\theta d\theta d\phi$ is the solid angle element. \emph{Fluctuating} variables are defined with respect to the Reynolds and Favre averages as
\begin{equation}
    X' = X - \hat{X}, \qquad X'' = X - \widetilde{X}.
\end{equation}

\section{Capturing the magnetorotational instability}
\label{sec:MRI_resol}

Our 3D simulations show an amplification of magnetic field strength, particularly in convective regions where the saturated field strength follows 
a shear-based saturation estimate formally similar to the MRI saturation estimate. To resolve magnetic instabilities in our simulations we need sufficient spatial resolution to capture the fastest growing mode along with sufficient simulation time to clearly see the impact of the instability on the magnetic field.

\begin{figure}[t!]
    \centering
    \includegraphics[width=\columnwidth]{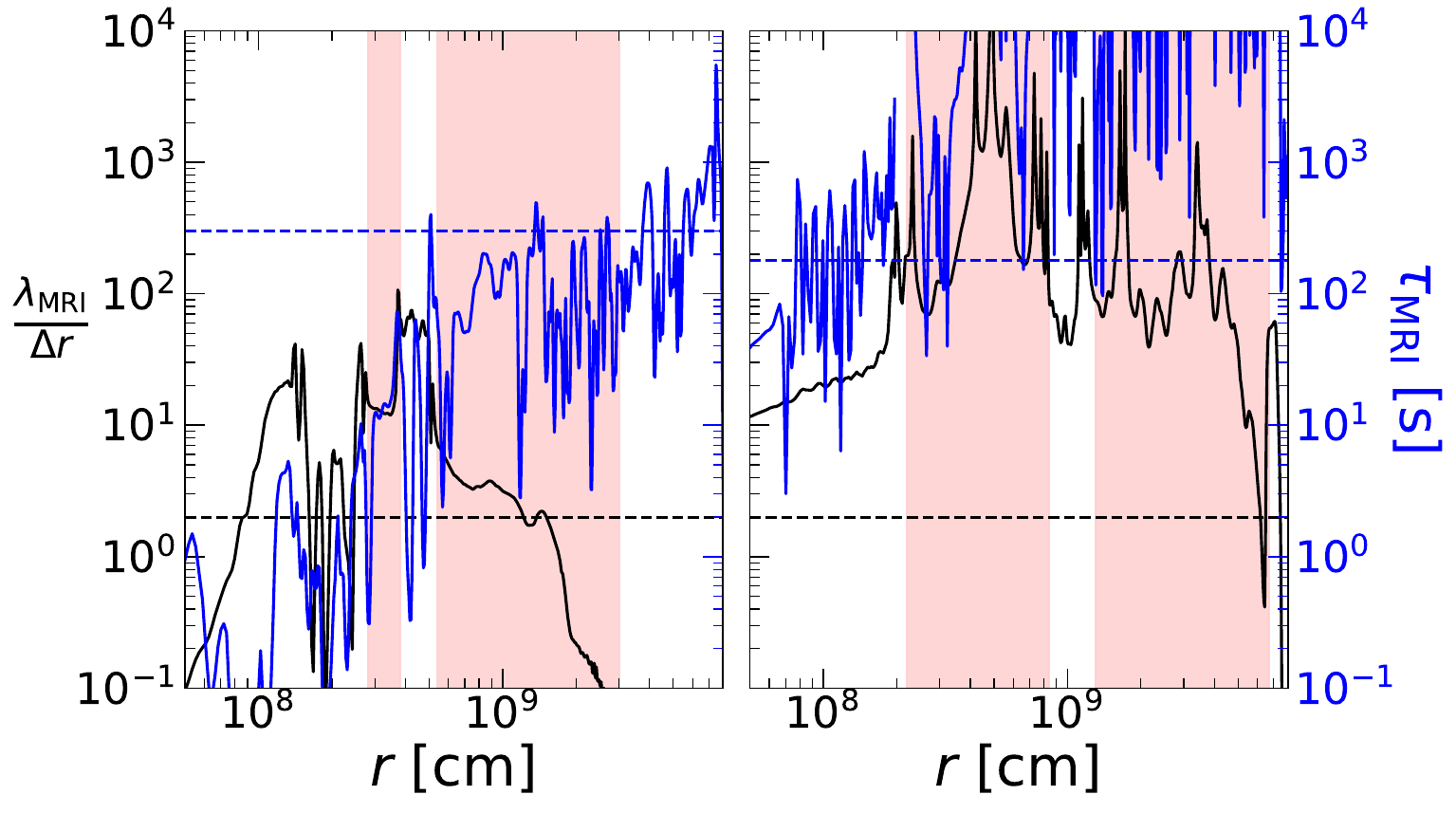}
    \caption{Ratio of the fastest growing mode, Eq.~\ref{eqn:fast_MRI}, and the radial grid size in model \ad (left) and model \ag (right). The dashed black line indicates $\lambda_{\rm MRI}/\Delta r = 2$. In blue we show both the growth timescale of the MRI, Eq.~\ref{eqn:time_MRI}, and the total 3D simulation time for each model (dashed horizontal line). The red zones delimit the convective regions for each model.
    }
    \label{fig:MRI_resol}
\end{figure}

For the MRI the fastest growing mode is roughly given by \citep{Balbus_Hawley_1991},
\begin{equation}
\label{eqn:fast_MRI}
    \lambda_{\rm MRI} = \frac{2 \pi B}{\Omega\sqrt{\rho}}.
\end{equation}
To capture MRI growth, the spatial resolution must be sufficient to resolve the fastest-growing mode, with at least two grid cells per wavelength, although larger quality factors are preferable. Capturing the nonlinear termination of the MRI by parasitic instabilities requires somewhat higher resolution, typically of order 2--4 times finer than the fastest-growing wavelength \citep[see][]{Rembiasz_2016}. Since the radial resolution of our grid is not uniform, Fig.~\ref{fig:MRI_resol} shows the ratio between the fastest-growing MRI wavelength and the radial cell size as a function of radius. In model \ag (right panel), this ratio is well above 10 throughout the computational domain, owing to the slow rotation of the model, which leads to large values of $\lambda_{\rm MRI}$. In model \ad, the fastest-growing wavelength is much smaller. Nevertheless, according to this resolution criterion, the MRI is resolved throughout the convective zone C1, the radiative zone R2, and the base of C2.

We further estimate the typical growth timescale of the MRI as,
\begin{equation}
\label{eqn:time_MRI}
    \tau_{\rm MRI} = \gamma^{-1}_{\rm MRI} = \left(\frac{q}{2}\Omega\right)^{-1},
\end{equation}
also shown in Fig.~\ref{fig:MRI_resol}. This timescale is compared to the total duration of the 3D simulation for each model. Due to the slow rotation of \ag the growth time of the MRI is longer than the simulated time in most of the simulated domain. For \ad the simulated time is long enough to capture multiple growth times in the regions where the MRI is spatially resolved, such as C1 and R2.

\section{Evolution of the magnetic field near convective boundaries}
\label{sec:magbnd_evol}

The evolution of the magnetic field at convective boundaries provides a useful diagnostic of magnetic-flux exchange between turbulent and radiative regions. This is particularly relevant for simulations that excise part of the stellar interior or impose magnetic boundary conditions, since the field strength near convective-shell boundaries can influence the Maxwell stresses and, consequently, the direction and efficiency of AM transport. In our global models, these boundary fields are not prescribed externally, but evolve self-consistently as the convective shells interact with their neighbouring radiative layers. Here, we examine this evolution to assess whether the magnetic field near the shell boundaries is governed mainly by contraction and flux conservation, or whether it is also affected by local amplification and magnetic exchange with the turbulent regions.

\begin{figure}[t!]
    \centering
    \includegraphics[width=\columnwidth]{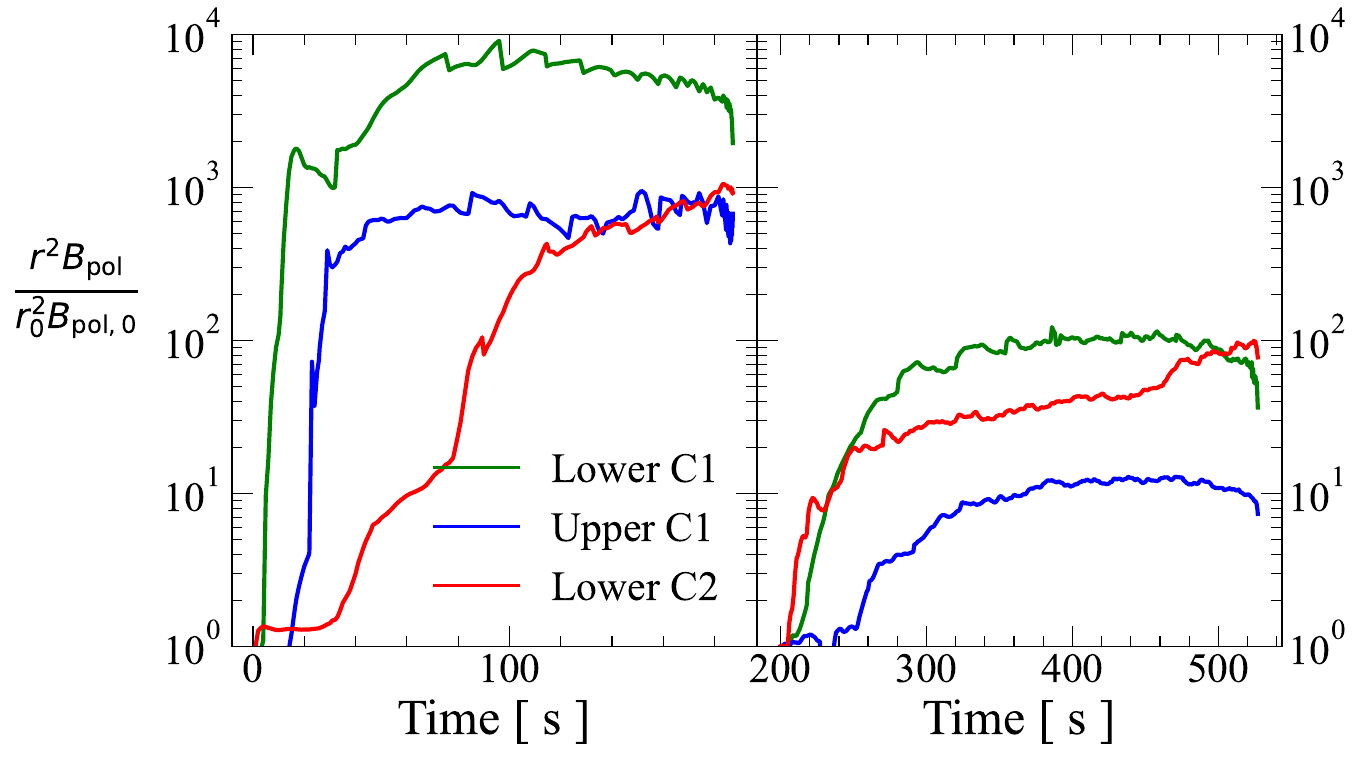}
    \caption{Evolution of the poloidal magnetic-field flux, normalised to its initial value, at selected convective boundaries for models \ag (left) and \ad (right). The three boundaries shown are the lower and upper boundaries of C1 and the lower boundary of C2, each tracked at fixed mass coordinate.}
    \label{fig:mag_boundaries_evol}
\end{figure}

%\maaC{Adam, I wonder whether it makes sense to also show the evolution of $(Br^2_{\rm b}) /(B_0r^2_{0,\rm b}) $, where $B$ and $B_0$ are the field strengths on the boundary at a given time (as represented in the figure above) and at the beginning of the simulation, and $r_{\rm b}$ and $r_{0,\rm b}$ the radius of the interface at the given time and at $t=0$. The idea is to show that the evolution is not triggered by flux conservation "only". If it were, $(Br^2_{\rm b}) /(B_0r^2_{0,\rm b}) \simeq constant$, but if there is local amplification, it won't be the case. I would add this as an appendix (already done), since it is a technical point, though important for models that aim at not simulating the core of the star. Stricty, this scaling works for the poloidal field. For the toroidal field, it is approximate (see text below). }
%\maaC{I we add the extra figure I have mentioned, we could add this paragraph:}

To separate passive compression from genuine field amplification, we monitor the flux-freezing scaling expected for the poloidal component. For a radial or poloidal field component, flux freezing through a spherical surface implies $B_r r^2\simeq\mathrm{const.}$.

%For a toroidal component, the conserved flux is instead the flux through a meridional material surface, so that, for a thin Lagrangian layer of thickness $\Delta r_b$,
%\[
%    B_{\phi,b} r_b \Delta r_b \simeq \mathrm{const.}
%\]
%Equivalently, for a fixed mass interval around the boundary, this may be written as $B_{\phi,b}/(\rho_b r_b)\simeq\mathrm{const.}$. Significant deviations from these scalings indicate that the boundary field is not evolving by compression alone, but is affected by winding, turbulent transport, local amplification, or magnetic exchange with the adjacent layers.
%\maaC{Note that if the contraction is homologous, $\Delta r \propto r$ and, hence, $ B_{\phi,b} r_b \Delta r_b \propto B_{\phi,b} r_b^2 \simeq \mathrm{const.}$. }

Figure~\ref{fig:mag_boundaries_evol} shows the evolution of the poloidal magnetic-field flux, normalised to its initial value, at selected convective boundaries. We find clear amplification of the poloidal field at all boundaries. The boundaries of C1 approach a steady state by the end of the simulation, whereas the lower boundary of C2 has not yet reached a settled state. This is consistent with the slower development of turbulence in C2, discussed in \citetalias{Griffiths2026PaperI}, which indicates that this region requires more convective turnovers to reach a statistically stationary state.

The growth of the boundary field is not simply a passive consequence of contraction. As shown in Fig.~\ref{fig:magnetic_field_start_end}, the magnetic field amplified inside convective regions interacts continuously with the surrounding radiative layers. Magnetic flux is transported across the convective--radiative interfaces, enhancing the field in the neighbouring radiative zones while also modifying the magnetic-field evolution inside the turbulent shells. This exchange may affect the saturation level reached in the convective regions and is therefore relevant for the resulting Maxwell stresses. Given the sensitivity of AM transport to magnetic-field strength, the self-consistent evolution of the field near convective boundaries can be important for determining the final AM profile of the progenitor.

\section{3D rendering of magnetic field lines}
\label{sec:3D_vis}
\begin{figure*}[ht!]
    \centering
    \includegraphics[width=0.75\textwidth]{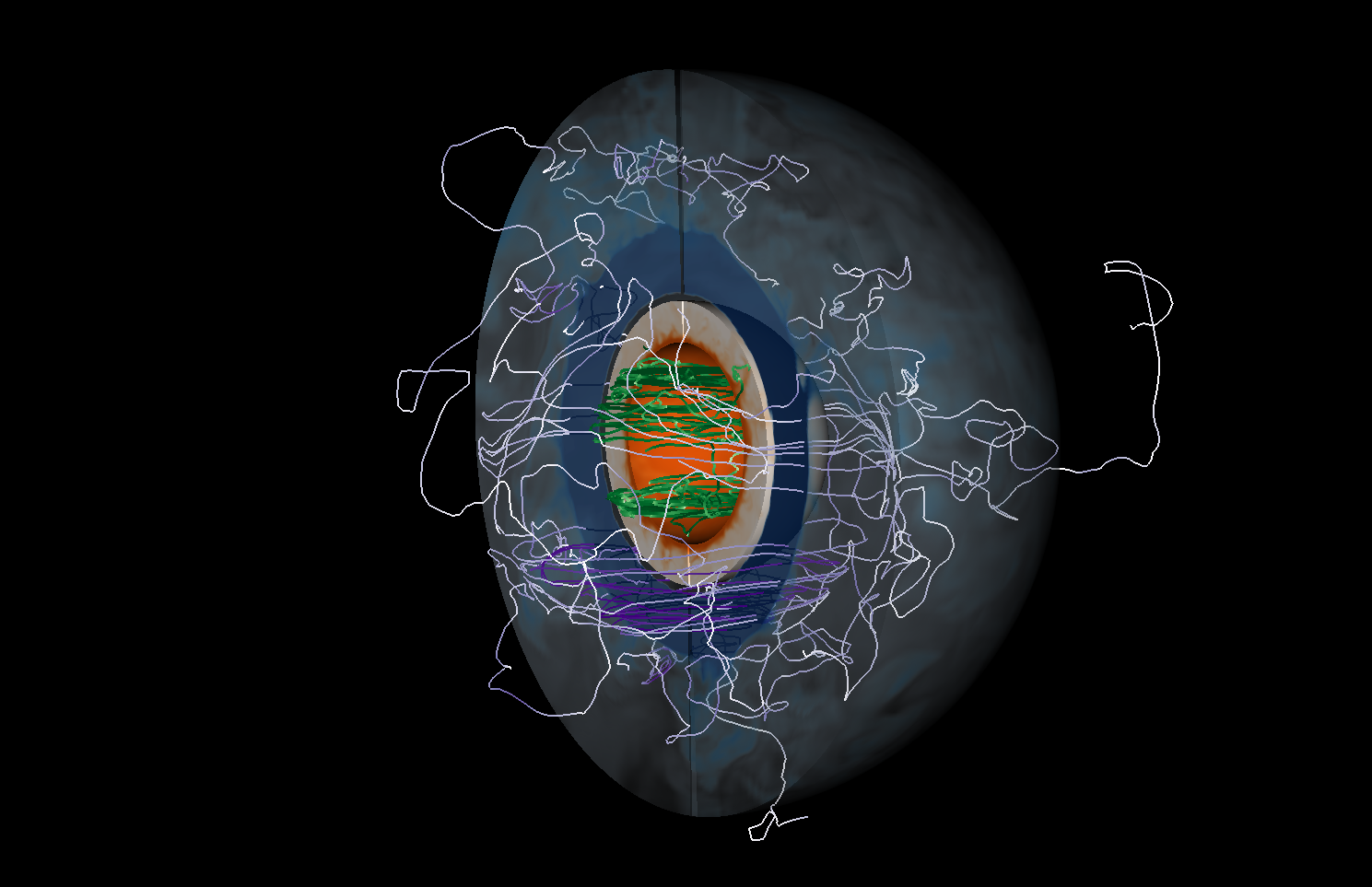}
    \includegraphics[width=0.75\textwidth]{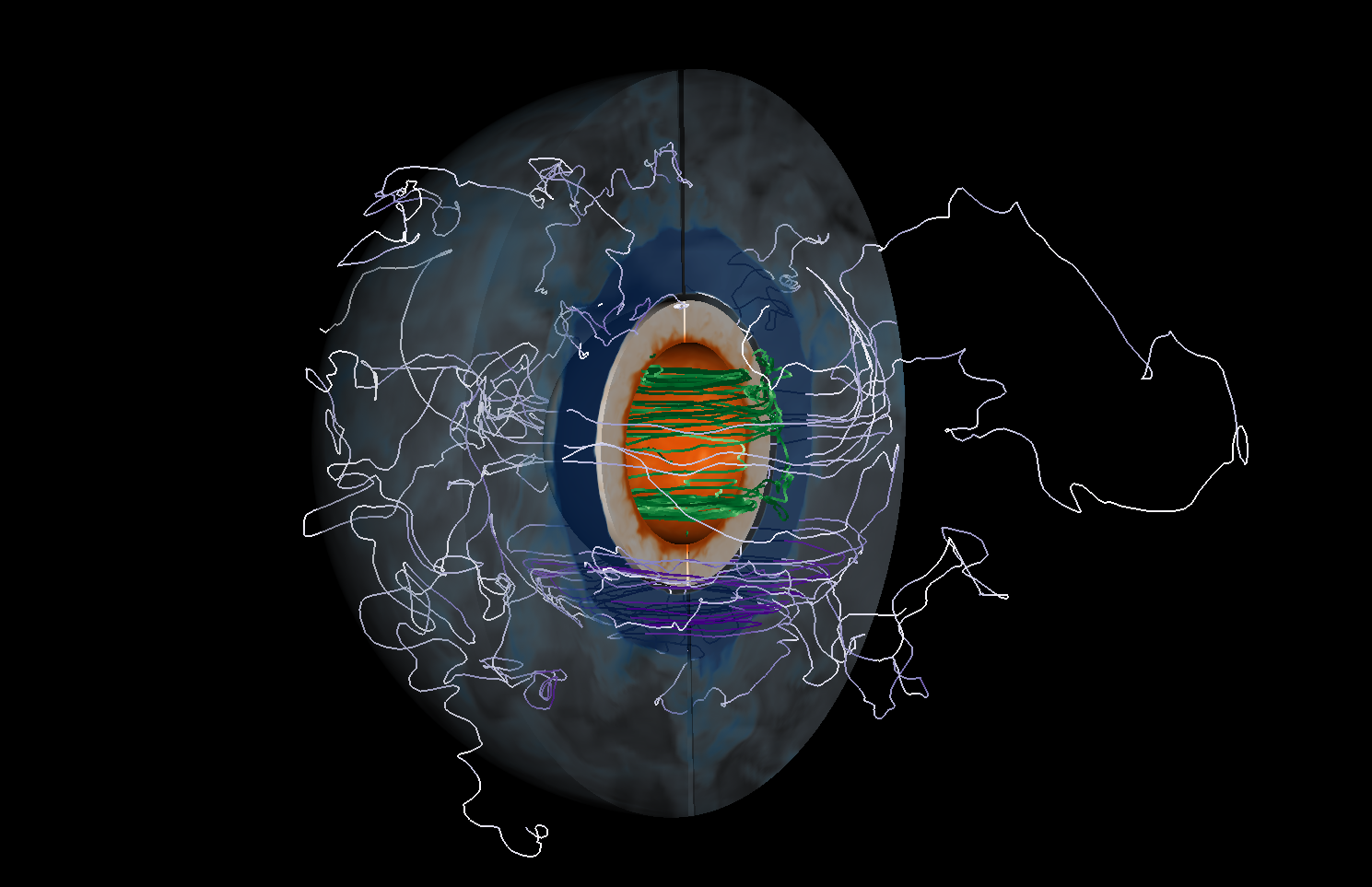}
    \caption{3D rendering of the inner oxygen shell (blue shading) and silicon-rich region (orange shading) of model \ad at collapse. Magnetic field lines are traced from foot points located in the oxygen layer (purple lines) and inside the radiative iron core (green lines). The two panels show different viewing angles: right side view above and left side view below.}
    \label{fig:snapshot_right}
\end{figure*}

We visualise the 3D structure of the magnetic field lines of model \ad in Fig.~\ref{fig:snapshot_right}.
Field lines rooted in the radiative core remain strongly wound around the rotation axis, consistent with the toroidal dominance found in the spectral analysis. Field lines rooted in the oxygen shell extend over a larger radial range and display a more tangled geometry, reflecting the stronger poloidal component and the multi-scale structure generated by turbulence. 

%MAA: Let's separate 

This topology implies that the collapsing core and the surrounding burning shells are not magnetically isolated once the multidimensional evolution is taken into account. Instead, magnetic flux transported and amplified in the convective layers can connect regions that were initially separated in the 1D magnetic-field prescription.

\end{appendix}
\end{document}